\documentclass[twoside]{article}
\usepackage{PRIMEarxiv}

\usepackage[utf8]{inputenc} 
\usepackage[T1]{fontenc}    
\usepackage{hyperref}       
\usepackage{url}            
\usepackage{booktabs}       
\usepackage{amsfonts}       
\usepackage{nicefrac}       
\usepackage{microtype}      
\usepackage{lipsum}
\usepackage{fancyhdr}       
\usepackage{graphicx}       
\graphicspath{{media/}}     
\usepackage{setspace}

\usepackage{cite}
\usepackage{amsmath,amssymb,amsthm,mathrsfs,amsfonts,dsfont} 
\DeclareMathAlphabet{\mathpzc}{OT1}{pzc}{m}{it}
\usepackage{algorithm,algorithmic,setspace}
\usepackage{graphicx}
\usepackage{textcomp}
\usepackage{xcolor}
\usepackage{tabularx,booktabs}
\usepackage{adjustbox}
\usepackage{array,etoolbox}
\usepackage{dcolumn}
\usepackage{rotating}
\usepackage{pifont}
\usepackage{tikz}
\usepackage{comment}
\usepackage{dsfont}
\usepackage{multicol}
\usepackage{multirow}
\usepackage{mathtools}
\usepackage{stfloats}
\usepackage{lettrine}

\usepackage[nopostdot,style=super,nonumberlist, toc]{glossaries}
\usepackage{tabularx,booktabs}
\usepackage[utf8]{inputenc}
\usepackage{fourier} 
\usepackage{array}
\usepackage{makecell}
\usepackage[nopostdot,style=super,toc]{glossaries}

\title{Device-to-Device Communication in 5G/6G: Architectural Foundations and Convergence with Enabling Technologies
\thanks{This work has been submitted to the IEEE for possible publication.  Copyright may be transferred without notice, after which this version may no longer be accessible.}
}

\author{
  Mohammad Reza Fasihi, Brian L. Mark \\
  Dept. of Electrical and Computer Engineering and Wireless Cyber Center \\
  George Mason University \\
  Fairfax, Virginia, United States\\
  \texttt{\{mfasihi4, bmark\}@gmu.edu} \\
}

\date{July 2025}

\makeglossaries
\newglossaryentry{FCC}{%
  name={FCC},
  description={Federal Communications Commission}
}
\newglossaryentry{RAT}{%
  name={RAT},
  description={Radio Access Technology}
}
\newglossaryentry{ISM}{%
  name={ISM},
  description={Industrial, Scientific, and Medical}
}
\newglossaryentry{Wi-Fi}{%
  name={Wi-Fi},
  description={Wireless Fidelity}
}
\newglossaryentry{WiGig}{%
  name={WiGig},
  description={Wireless Gigabit}
}
\newglossaryentry{3GPP}{%
  name={3GPP},
  description={3rd Generation Partnership Project}
}
\newglossaryentry{AR}{%
  name={AR},
  description={Augmented Reality}
}
\newglossaryentry{VR}{%
  name={VR},
  description={Virtual Reality}
}
\newglossaryentry{CCI}{%
  name={CCI},
  description={Co-Channel Interference}
}
\newglossaryentry{LBT}{%
  name={LBT},
  description={Listen Before Talk}
}
\newglossaryentry{CCA}{%
  name={CCA},
  description={Clear Channel Assessment}
}
\newglossaryentry{LTE}{%
  name={LTE},
  description={Long Term Evolution}
}
\newglossaryentry{LAA}{%
  name={LAA},
  description={Licensed-Assisted Access}
}
\newglossaryentry{LTE-U}{%
  name={LTE-U},
  description={LTE Unlicensed}
}
\newglossaryentry{NR}{%
  name={NR},
  description={New Radio}
}
\newglossaryentry{NR-U}{%
  name={NR-U},
  description={New Radio Unlicensed}
}
\newglossaryentry{4G}{%
  name={4G},
  description={4th Generation}
}
\newglossaryentry{5G}{%
  name={5G},
  description={5th Generation}
}
\newglossaryentry{6G}{%
  name={6G},
  description={6th Generation}
}
\newglossaryentry{mmWave}{%
  name={mmWave},
  description={Millimeter-Wave}
}
\newglossaryentry{CA}{%
  name={CA},
  description={Carrier Aggregation}
}
\newglossaryentry{U-NII}{%
  name={U-NII},
  description={Unlicensed National Information Infrastructure}
}
\newglossaryentry{EC}{%
  name={EC},
  description={European Commission}
}
\newglossaryentry{CSMA/CA}{%
  name={CSMA/CA},
  description={Carrier-Sense Multiple Access with Collision Avoidance}
}
\newglossaryentry{AP}{%
  name={AP},
  description={Access Point}
}
\newglossaryentry{CHS}{%
  name={CHS},
  description={Channel Selection}
}
\newglossaryentry{CSAT}{%
  name={CSAT},
  description={Carrier Sense Adaptive Transmission}
}
\newglossaryentry{ABS}{%
  name={ABS},
  description={Almost Blank Subframe}
}
\newglossaryentry{FBS}{%
  name={FBS},
  description={Fully Blank Subframe}
}
\newglossaryentry{DRX}{%
  name={DRX},
  description={Discontinuous Reception}
}
\newglossaryentry{TPC}{%
  name={TPC},
  description={Transmit Power Control}
}
\newglossaryentry{IoT}{%
  name={IoT},
  description={Internet of Things}
}
\newglossaryentry{SL}{%
  name={SL},
  description={Sidelink}
}
\newglossaryentry{D2D}{%
  name={D2D},
  description={Device-to-Device}
}
\newglossaryentry{SL-U}{%
  name={SL-U},
  description={Sidelink in Unlicensed Spectrum}
}
\newglossaryentry{C-V2X}{%
  name={C-V2X},
  description={Cellular Vehicle-to-Everything}
}
\newglossaryentry{D2D-U}{%
  name={D2D-U},
  description={D2D in unlicensed Spectrum}
}
\newglossaryentry{EE}{%
  name={EE},
  description={Energy Efficiency}
}
\newglossaryentry{SE}{%
  name={SE},
  description={Spectrum Efficiency}
}
\newglossaryentry{HARQ}{%
  name={HARQ},
  description={Hybrid Automatic Repeat Request}
}
\newglossaryentry{ProSe}{%
  name={ProSe},
  description={Proximity Services}
}
\newglossaryentry{CAM}{%
  name={CAM},
  description={Cooperative Awareness Message}
}
\newglossaryentry{DENM}{%
  name={DENM},
  description={Decentralized Environmental Notification Message}
}
\newglossaryentry{QAM}{%
  name={QAM},
  description={quadrature amplitude modulation}
}
\newglossaryentry{LTE-A}{%
  name={LTE-A},
  description={LTE-Advanced}
}
\newglossaryentry{IIoT}{%
  name={IIoT},
  description={Industrial IoT}
}
\newglossaryentry{IBD}{%
  name={IBD},
  description={In-Band D2D}
}
\newglossaryentry{OBD}{%
  name={OBD},
  description={Out-Band D2D}
}
\newglossaryentry{M2M}{%
  name={M2M},
  description={Machine-to-Machine}
}
\newglossaryentry{RRM}{%
  name={RRM},
  description={Radio Resource Management}
}
\newglossaryentry{MIMO}{%
  name={MIMO},
  description={Multiple-Input Multiple-Output}
}
\newglossaryentry{MISO}{%
  name={MISO},
  description={Multiple-Input Single-Output}
}
\newglossaryentry{WiMAX}{%
  name={WiMAX},
  description={Worldwide Interoperability for Microwave Access}
}
\newglossaryentry{MU-MIMO}{%
  name={MU-MIMO},
  description={Multi-User MIMO}
}
\newglossaryentry{mMIMO}{%
  name={mMIMO},
  description={Massive MIMO}
}
\newglossaryentry{OMA}{%
  name={OMA},
  description={Orthogonal Multiple Access}
}
\newglossaryentry{NOMA}{%
  name={NOMA},
  description={Non-Orthogonal Multiple Access}
}
\newglossaryentry{RSMA}{%
  name={RSMA},
  description={Rate-Splitting Multiple Access}
}
\newglossaryentry{DoF}{%
  name={DoF},
  description={Degree of Freedom}
}
\newglossaryentry{SNR}{%
  name={SNR},
  description={Signal-to-Noise Ratio}
}
\newglossaryentry{ASR}{%
  name={ASR},
  description={Average Sum Rate}
}
\newglossaryentry{ML}{%
  name={ML},
  description={Machine Learning}
}
\newglossaryentry{DL}{%
  name={DL},
  description={Deep Learning}
}
\newglossaryentry{RL}{%
  name={RL},
  description={Reinforcement Learning}
}
\newglossaryentry{DRL}{%
  name={DRL},
  description={Deep Reinforcement Learning}
}
\newglossaryentry{DQN}{
  name={DQN},
  description={Deep Q Network}
}
\newglossaryentry{DDPG}{
  name={DDPG},
  description={Deep Deterministic Policy Gradient}
}
\newglossaryentry{TDD}{%
  name={TDD},
  description={Time Division Duplex}
}
\newglossaryentry{FDD}{%
  name={FDD},
  description={Frequency Division Duplex}
}
\newglossaryentry{PPP}{%
  name={PPP},
  description={Poisson Point Process}
}
\newglossaryentry{PCP}{%
  name={PCP},
  description={Poisson Cluster Process}
}
\newglossaryentry{JSDM}{%
  name={JSDM},
  description={Join Spatial Division and Multiplexing}
}
\newglossaryentry{CS}{%
  name={CS},
  description={Compressive Sensing}
}
\newglossaryentry{MSE}{%
  name={MSE},
  description={Mean Squared Error}
}
\newglossaryentry{DAC}{%
  name={DAC},
  description={Digital-to-Analog Converter}
}
\newglossaryentry{RF}{%
  name={RF},
  description={Radio Frequency}
}
\newglossaryentry{SWIPT}{%
  name={SWIPT},
  description={Simultaneous Wireless Information and Power Transfer}
}
\newglossaryentry{RIS}{%
  name={RIS},
  description={Reconfigurable Intelligent Surface}
}
\newglossaryentry{CR}{%
  name={CR},
  description={Cognitive Radio}
}

\newglossaryentry{HD}{%
  name={HD},
  description={Half-Duplex}
}

\newglossaryentry{FD}{%
  name={FD},
  description={Full-Duplex}
}

\newglossaryentry{SI}{%
  name={SI},
  description={Self Interference}
}
\newglossaryentry{LoS}{%
  name={LoS},
  description={Line-of-Sight}
}
\newglossaryentry{NLoS}{%
  name={NLoS},
  description={Non-LoS}
}
\newglossaryentry{RSI}{%
  name={RSI},
  description={Residual Self Interference}
}
\newglossaryentry{HetNet}{%
  name={HetNet},
  description={Heterogeneous Networks}
}
\newglossaryentry{GNN}{%
  name={GNN},
  description={Graph Neural Network}
}
\newglossaryentry{RB}{%
  name={RB},
  description={Resource Block}
}
\newglossaryentry{FDMA}{%
  name={FDMA},
  description={Frequency Domain Multiple Access}
}
\newglossaryentry{TDMA}{%
  name={TDMA},
  description={Time Domain Multiple Access}
}
\newglossaryentry{OFDMA}{%
  name={OFDMA},
  description={Orthogonal Frequency Division Multiple Access}
}
\newglossaryentry{SC}{%
  name={SC},
  description={Superposition Coding}
}
\newglossaryentry{SIC}{%
  name={SIC},
  description={Successive Interference Cancellation}
}
\newglossaryentry{URLLC}{%
  name={URLLC},
  description={Ultra-Reliable Low Latency Communications}
}
\newglossaryentry{EH}{%
  name={EH},
  description={Energy Harvesting}
}
\newglossaryentry{LBS}{%
  name={LBS},
  description={Location-Based Services}
}
\newglossaryentry{WLAN}{%
  name={WLAN},
  description={Wireless Local Area Networks}
}
\newglossaryentry{GNSS}{%
  name={GNSS},
  description={Global Navigation Satellite System}
}
\newglossaryentry{AoA}{%
  name={AoA},
  description={Angle of Arrival}
}
\newglossaryentry{POMDP}{%
  name={POMDP},
  description={Partially Observed Markov Decision Process}
}
\newglossaryentry{SAGIN}{%
  name={SAGIN},
  description={Space-Air-Ground Integrated Networks}
}
\newglossaryentry{NTN}{%
  name={NTN},
  description={Non-Terrestrial Network}
}
\newglossaryentry{ITS}{
  name={ITS},
  description={Intelligent Transportation System}
}
\newglossaryentry{LTE-V2X}{
  name={LTE-V2X},
  description={LTE-Based Vehicle-to-Everything}
}
\newglossaryentry{UAV}{
  name={UAV},
  description={Unmanned Aerial Vehicles}
}
\newglossaryentry{XR}{
  name={XR},
  description={Extended Reality}
}
\newglossaryentry{RedCap}{
  name={RedCap},
  description={Reduced Capability}
}
\newglossaryentry{MBMS}{
  name={MBMS},
  description={Multicast and Broadcast}
}
\newglossaryentry{BS}{
  name={BS},
  description={Base Station}
}
\newglossaryentry{CSI}{
  name={CSI},
  description={Channel State Information}
}
\newglossaryentry{QoS}{
  name={QoS},
  description={Quality of Service}
}
\newglossaryentry{IA}{
  name={IA},
  description={Interference Alignment}
}
\newglossaryentry{SINR}{
  name={SINR},
  description={Signal-to-Interference-Plus-Noise Ratio}
}
\newglossaryentry{FL}{
  name={FL},
  description={Federated Learning}
}
\newglossaryentry{MA}{
  name={MA},
  description={Multiple Access}
}
\newglossaryentry{SCMA}{
  name={SCMA},
  description={Sparse Code Multiple Access}
}
\newglossaryentry{PDMA}{
  name={PDMA},
  description={Pattern Division Multiple Access}
}
\newglossaryentry{ISAC}{
  name={ISAC},
  description={Integrated Sensing And Communication}
}
\newglossaryentry{GEO}{
  name={GEO},
  description={Geostationary Earth Orbit}
}
\newglossaryentry{MEO}{
  name={MEO},
  description={Medium Earth Orbit}
}
\newglossaryentry{LEO}{
  name={LEO},
  description={Low Earth Orbit}
}
\newglossaryentry{UE}{
  name={UE},
  description={User Equipment}
}

\begin{document}

\maketitle

\begin{abstract}

Device-to-Device (D2D) communication is a promising solution to meet the growing demands of 5G and future 6G networks by enabling direct communication between user devices. It enhances spectral efficiency (SE) and energy efficiency (EE), reduces latency, and supports proximity-based services. As wireless systems evolve toward 5G and 6G paradigms, the integration of D2D with advanced cellular technologies introduces new opportunities and challenges. This survey paper reviews the architectural foundations of D2D communication and explores its integration with key 5G/6G enabling technologies. We review standardization efforts, analyze core challenges, and highlight future research directions to unlock the full potential of D2D in next-generation wireless networks.
\end{abstract}


\section{INTRODUCTION}
\label{sec:introduction}

\lettrine[findent=2pt]{\textbf{T}} {he rapid proliferation} of smart devices, high-definition content, and real-time applications has placed unprecedented demands on wireless networks. As we move toward the fifth and sixth generations of cellular technologies (\gls{5G} and \gls{6G}), Device-to-Device (\gls{D2D}) communication has emerged as a key enabler to meet these growing demands. By allowing user equipments (\gls{UE}s) to communicate directly without routing data through the base station (BS), D2D communication can improve spectral efficiency (\gls{SE}) and energy efficiency (\gls{EE}), reduce end-to-end latency, and support new proximity-based services. These capabilities make D2D a promising solution for a wide range of emerging applications such as vehicular communications, public safety, and massive Internet of Things (\gls{IoT}) deployments. 

At the same time, D2D communication introduces a range of technical challenges, particularly when deployed alongside evolving cellular architectures. Issues such as interference management, mode selection, dynamic resource allocation, and mobility support must be addressed to fully realize its potential. These challenges become even more complex when D2D is integrated with advanced technologies that define the 5G/6G landscape. For example, massive Multiple Input Multiple Output (\gls{MIMO}) and millimeter-wave (\gls{mmWave}) systems offer high capacity and data rates but require careful spatial coordination when D2D links are active. Similarly, novel multiple access techniques such as Non-Orthogonal Multiple Access (\gls{NOMA}) and Rate-Splitting Multiple Access (\gls{RSMA}) present new opportunities for optimizing D2D interactions, albeit with added complexity. Localization technologies are also playing an increasing role in enabling intelligent and context-aware D2D connections. Additionally, D2D systems are being explored in unlicensed spectrum bands like those accessed via Licensed Assisted Access (\gls{LAA}) or 5G NR-U, which raises the need for efficient coexistence with incumbents such as Wi-Fi. In parallel, the emergence of integrated network architectures like Satellite-Air-Ground Integrated Networks (\gls{SAGIN}) and Integrated Sensing and Communication (\gls{ISAC}) further expands the scope of D2D by enabling extended coverage, resilience, and connectivity in remote and dynamic environments.

This survey aims to provide a structured and in-depth review of the evolving role of D2D communication within the broader context of 5G/6G systems. Instead of treating each enabling technology in isolation, we adopt a challenge-oriented perspective that examines how D2D interacts with key technologies through the lenses of core design issues such as mode selection, channel estimation, resource allocation, and interference mitigation. The main contributions of this survey are summarized as follows:
\begin{itemize}
    \item We review the current status of D2D-related standardization, focusing on 3GPP efforts such as Proximity Services (\gls{ProSe}), the 5G NR sidelink interface, and ongoing developments relevant to 6G networks.
    \item We provide a comprehensive introduction to the basic concepts of D2D in cellular networks, including in-band and out-band operation, relay-based communication, and full-duplex transmission modes.
    \item We present a detailed literature review on the integration of D2D with enabling 5G/6G technologies such as massive MIMO, mmWave communication, advanced multiple access schemes, localization, unlicensed spectrum sharing, SAGIN, and ISAC.
\end{itemize}

The remainder of this paper is organized as follows. Section~\ref{sec:standardization} reviews standardization efforts surrounding D2D communication, focusing on 3GPP ProSe, 5G NR sidelink, and recent developments targeting 6G. Section~\ref{sec:d2d_fundamentals_in_cellular} introduces the fundamental concepts of D2D communication in cellular systems, covering key classifications and operational modes. Section~\ref{sec:d2d_5G_interplay} discusses the general integration of D2D within cellular architectures, highlighting design considerations and deployment models. Section~\ref{sec:d2d_massive_mimo} through~\ref{sec:d2d_multiple_access} explore the interplay between D2D and major enabling technologies: massive MIMO, mmWave communications, and multiple access mechanisms, respectively. Section~\ref{sec:d2d_misc} covers additional emerging domains, including localization, unlicensed spectrum use, SAGIN, and ISAC. Section~\ref{sec:future_research} outlines key open challenges and future research directions. Finally, Section~\ref{sec:conclusions} concludes the paper.

\section{Overview of D2D Communication and Its Evolution in 3GPP Standards}
\label{sec:standardization}

\begin{figure*}[t]
	\includegraphics[width=\linewidth]{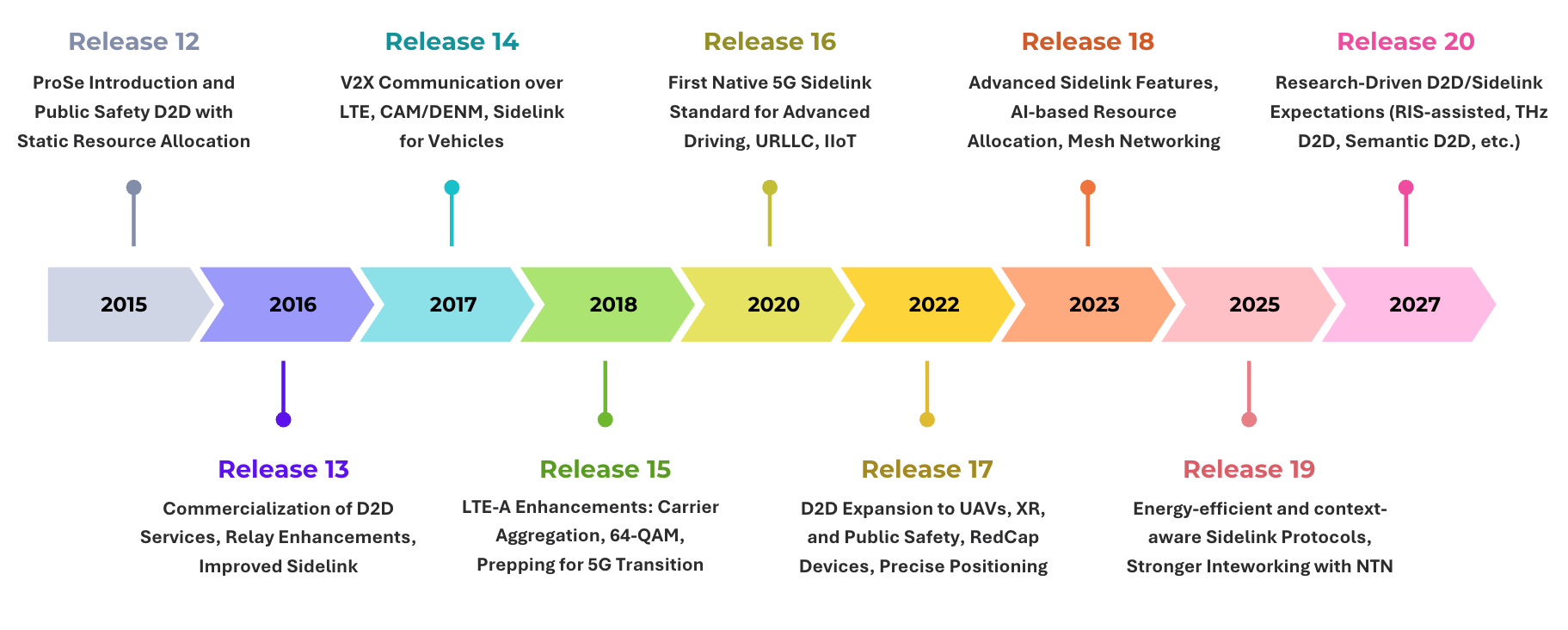}
	\caption{Evolution of D2D communication across 3GPP standards (Release 12 to Release 20).}
	\label{fig:3GPP_D2D_timeline}
\end{figure*}

D2D communication is a wireless communication model where user devices exchange data directly without involving the base station (\gls{BS}) or the core network. In such scenarios, two UE devices establish a sidelink connection, enabling direct communication when they are within close range. The sidelink defines the structure of logical, transport, and physical channels that support D2D connectivity, and the necessary radio resources are typically allocated by the cellular operator from a predefined resource pool~\cite{Schlienz:2015}. 

This proximity-based communication approach offers several key advantages. By bypassing the central network infrastructure, D2D reduces transmission delays, enhances energy efficiency (\gls{EE}) and spectral efficiency (\gls{SE}), and relieves congestion on the core network. It is particularly useful in applications like public safety communications, V2X, local content sharing, and social or proximity-based services. D2D can operate in either in-band mode—sharing spectrum with cellular services—or out-band mode using unlicensed spectrum technologies such as Wi-Fi or Bluetooth~\cite{Liu:2015}. Its flexible architecture and relatively simple setup also make it easy to integrate with existing wireless technologies to boost overall network performance. Furthermore, when the operator has knowledge of UE locations, it can intelligently manage radio resource allocation and interference coordination between D2D and conventional cellular users. Fig.~\ref{fig:3GPP_D2D_timeline} illustrates the evolution of D2D communication across 3GPP standards.

\subsection{ProSe Introduction and Public Safety D2D}

The standardization of D2D communication began with 3GPP Release 12, introduced under the name ProSe, primarily motivated by public safety requirements in scenarios where cellular network coverage could not be guaranteed~\cite{Schlienz:2015}. This release defined three modes for establishing D2D links: in-coverage, out-of-coverage, and partial coverage. In in-coverage mode, the cellular network allocates radio resources to transmitting UEs from a predefined pool. In contrast, out-of-coverage users rely on preconfigured settings for resource allocation, as network assistance is unavailable. Partial coverage users may operate in either mode, depending on their connectivity to the BS~\cite{Schlienz:2015}. 

Release 12 also introduced foundational features such as discovery and direct communication between UEs, supporting both one-to-one and one-to-many transmission modes over the sidelink interface. These features were particularly targeted at critical and public safety scenarios. However, the release had notable limitations—it supported only static resource allocation and lacked support for feedback-based transmission schemes such as link adaptation and Hybrid Automatic Repeat Request (\gls{HARQ}). Additionally, UE-to-network relay functionality was not yet included in this initial standardization effort.

\subsection{Commercialization and Relay Enhancements}

Building upon Release 12, 3GPP Release 13 extended ProSe features to support commercial D2D services and introduced enhancement in sidelink communication, such as improved discovery mechanisms and resource allocation procedures. It also added support for relay functionality, allowing an in-coverage UE to forward data to out-of-coverage UEs, which is particularly useful in coverage extension and disaster recovery situations~\cite{Lien:2020}. Moreover, sidelink enhancements were designed to better support out-of-coverage scenarios where centralized scheduling is not possible.

\subsection{LTE-V2X and the Start of Vehicular D2D}

With Release 14, the scope of D2D communication was significantly broadened to encompass vehicular communication as part of the broader intelligent transportation system (\gls{ITS}) vision. This release marked the introduction of LTE-based vehicle-to-everything (\gls{LTE-V2X}), leveraging sidelink enhancements to support both network-scheduled and autonomous scheduling modes of D2D operation. A key innovation was the exchange of critical safety messages, namely the cooperative awareness message (\gls{CAM}) and the decentralized environmental notification message (DENM) among vehicles, pedestrians, and roadside infrastructure. These messages are essential for enabling cooperative driving behavior and situational awareness, laying the groundwork for the development of Cellular V2X (\gls{C-V2X}). Importantly, Release 14 represented the first standardized version of C-V2X based on the 4G LTE air interface. It also introduced channel sensing capabilities, allowing UEs to detect ongoing transmissions and avoid resource collisions, thereby improving communication reliability in highly dynamic vehicular environments~\cite{Gonzalez:2019}. 

\subsection{5G NR-Based Sidelink and C-V2X Standardization}

3GPP Releases 15 and 16 represent a pivotal transition from LTE-based sidelink to a more advanced framework under the 5G New Radio (\gls{NR}) architecture. While maintaining backward compatibility with existing LTE sidelink protocols, these releases introduced substantial enhancements to support the emerging requirements of 5G-enabled D2D communication. Specifically, Release 15 focused on improving the performance of LTE-based sidelink by introducing features such as carrier aggregation (\gls{CA}), transmission diversity, and 64-QAM modulation, which collectively aimed to boost throughput and reduce transmission latency~\cite{Fodor:2019}. However, Releases 12 through 15 were still grounded in the LTE-Advanced (\gls{LTE-A}) air interface and lacked full integration with 5G network capabilities, particularly with respect to native sidelink transmission.

It was with Release 16 that 3GPP formally established a new sidelink standard based on 5G NR, specifically targeting C-V2X applications~\cite{Garcia:2021}. This release introduced NR-based sidelink communication with significant improvements in latency, reliability, scalability, and SE which are critical for real-time services, industrial automation, and advanced vehicular applications. The scope of Release 16 extended to both in-coverage and out-of-coverage scenarios, supporting features such as unicast, groupcast, and broadcast communication modes over the sidelink. To further reduce latency and support high connection density, grant-free transmission, previously used in NR uplink, was adopted for NR sidelink. Additionally, enhanced channel sensing, resource selection, congestion control, and QoS management mechanisms were introduced to mitigate resource collisions and ensure robust communication among densely deployed UEs~\cite{Lien:2020}. These innovations laid the groundwork for advanced driving use cases such as vehicle platooning, extended sensor sharing, and remote driving, marking a critical step toward the realization of next-generation cooperative and autonomous vehicular systems.

\subsection{D2D for UAVs, Extended Reality, and Public Safety}

Continuing with Release 17, 3GPP substantially expanded the scope of NR sidelink communication beyond traditional V2X use cases to support a broader range of D2D applications, including unmanned aerial vehicles (\gls{UAV}s), industrial automation, extended reality, and public safety communication in non-line-of-sight (\gls{NLoS}) conditions. This release introduced key enhancements such as dynamic resource allocation, improved power control, and support for unicast and multicast sidelink transmission. One of the notable advancements was the inclusion of reduced capability sidelink devices and network-controlled sidelink relaying, which together enhance the flexibility, scalability, and energy efficiency of D2D communication across diverse verticals.

To better serve mission-critical and public safety applications, Release 17 also incorporated enhancements to the ProSe framework. These include power-saving features for handheld and pedestrian devices, which are essential in emergency response scenarios and pedestrian safety use cases, such as direct communication between smartphones and vehicles in urban environments. The release also introduced support for pedestrian UEs within the V2X ecosystem and laid the groundwork for high-accuracy, low-latency positioning, which
is particularly valuable in industrial IoT (\gls{IIoT}) environments~\cite{Nokia:2020}.

\begin{figure*}
 \centering
	\includegraphics[width=\linewidth]{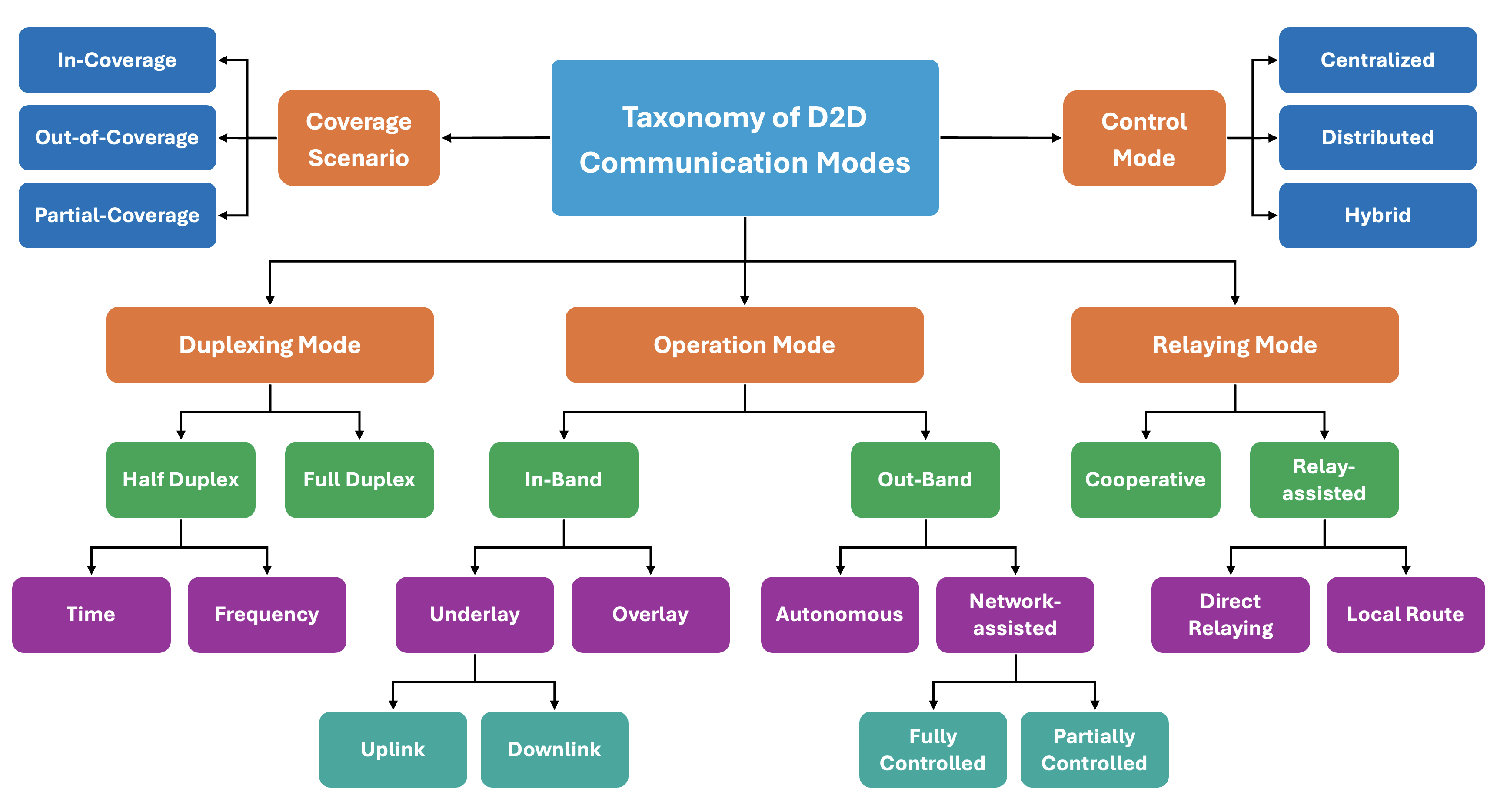}
	\caption{Taxonomy of D2D communication in cellular networks.}
	\label{fig:d2d_taxonomy}
\end{figure*}

\subsection{AI-Driven Sidelink and Multi-Hop Relaying}

The most recent 3GPP Release 18, developed under the umbrella of the 5G-Advanced initiative, continues to advance sidelink communication with a focus on enabling more sophisticated and high-density D2D applications. This release introduces a range of enhancements, including high-accuracy sidelink positioning, support for NR multicast and broadcast (\gls{MBMS}) over sidelink, and deeper integration of sidelink capabilities with emerging technologies such as network slicing and edge intelligence. These features are designed to support demanding use cases such as cooperative vehicular perception, vehicle platooning, industrial D2D, and other critical applications that require ultra-reliable low-latency communication (\gls{URLLC}).

One of the most notable advancements in Release 18 is the exploration of AI-assisted resource allocation for sidelink, which introduces a degree of adaptability and predictive intelligence in managing spectrum resources. Additionally, the release formalizes the concept of multi-hop sidelink relaying, a major step toward enabling mesh-like D2D networks capable of supporting complex and large-scale device deployments~\cite{WILEY_Jadav:2022}. These developments reflect the ongoing evolution of D2D communication within the 3GPP framework, positioning it as a foundational technology for 5G-Advanced and future 6G ecosystems.

\subsection{Toward Intelligent and 6G-Ready Sidelink Networking}

Looking ahead, 3GPP Release 19 and Release 20 are expected to further evolve Device-to-Device (D2D) communication as the wireless ecosystem moves toward 5G-Advanced and 6G. Release 19, still under development, is anticipated to enhance intelligent sidelink communication by embedding native AI/ML-based resource management, enabling context-aware and predictive scheduling to support dynamic environments such as industrial automation, UAV swarms, and dense urban V2X deployments. Additional improvements are expected in multi-hop relaying, energy-efficient sidelink protocols for reduced capability and battery-limited devices, and deeper interworking with non-terrestrial networks (\gls{NTN}s) to support D2D scenarios in remote or disaster-prone areas~\cite{Giuliano:2024}.

Building on these foundations, Release 20 will likely initiate the first standardization steps for 6G-oriented D2D networking, positioning sidelink as a primary communication mode in distributed, infrastructure-less environments. It is expected to introduce support for Reconfigurable Intelligent Surfaces (\gls{RIS}) to dynamically shape propagation paths for D2D links, and explore Terahertz D2D for ultra-high-throughput short-range communication. Further, concepts such as semantic and goal-oriented D2D exchange, native integration of sensing and positioning, and swarm intelligence for autonomous edge devices are expected to emerge. Together, these advancements signal a shift towards intelligent, cooperative, and context-adaptive D2D systems, laying the groundwork for next-generation applications in extended reality, smart factories, connected robotics, and autonomous mobility~\cite{Giuliano:2024}.

\section{Fundamentals of D2D Communication}
\label{sec:d2d_fundamentals_in_cellular}

D2D communication enables direct data exchange between nearby user devices without traversing the base station, offering potential gains in spectral efficiency, latency, and network offloading. Before diving into the technical synergy between D2D and emerging 5G/6G technologies, this section outlines the fundamental operational modes of D2D communication in cellular networks. We classify D2D by spectrum usage (in-band vs. out-band), relay involvement, duplexing capability, and control architecture, providing a foundation for the advanced interactions discussed in the subsequent sections.

The initial implementation of D2D communication in cellular networks aimed to facilitate multi-hop relay transmission. Subsequently, a diverse range of use cases emerged, including multicasting, peer-to-peer communication, content distribution, machine-to-machine (\gls{M2M}) and V2X communications, cellular offloading, and IoT applications~\cite{Asadi:2014}. The inherent flexibility and distinct advantages of D2D communication have established it as a pivotal element in enhancing cellular network performance. In this context, D2D communication is defined as the direct interaction between two mobile users without routing through the base station or core network. This mode of communication offers several benefits, including higher transmission rates, improved energy efficiency, alleviation of cellular traffic congestion, expanded network coverage, reduced latency, and enhanced spectral efficiency~\cite{Liu:2015}.

D2D communication promises three primary gains: proximity, hop, and reuse. The proximity gain results from the closeness of the communicating devices, ensuring high data rates, low energy consumption, and minimized delay. The reuse gain arises from the concurrent utilization of cellular spectrum by both cellular and D2D links, while the hop gain is achieved by replacing the conventional two-hop cellular transmission (uplink and downlink) with a single D2D link~\cite{Shaikh:2018}.

D2D communication is typically non-transparent to the cellular network and can be conducted over either licensed (i.e., cellular) spectrum, referred to as in-band, or unlicensed spectrum, known as out-band communication\cite{Gismalla:2022}. Additionally, D2D communication may operate in relay mode, where a device within coverage acts as an intermediary to forward data to devices experiencing poor connectivity or located outside the network coverage area. A graphical overview of this classification is 
presented in Fig.~\ref{fig:d2d_taxonomy}, which shows the taxonomy of D2D communication modes within cellular networks.

\subsection{Operation Modes of D2D Communications}
\label{subsec:operation_mode_d2d}

\begin{figure}
\centering
\includegraphics[width=0.8\linewidth]{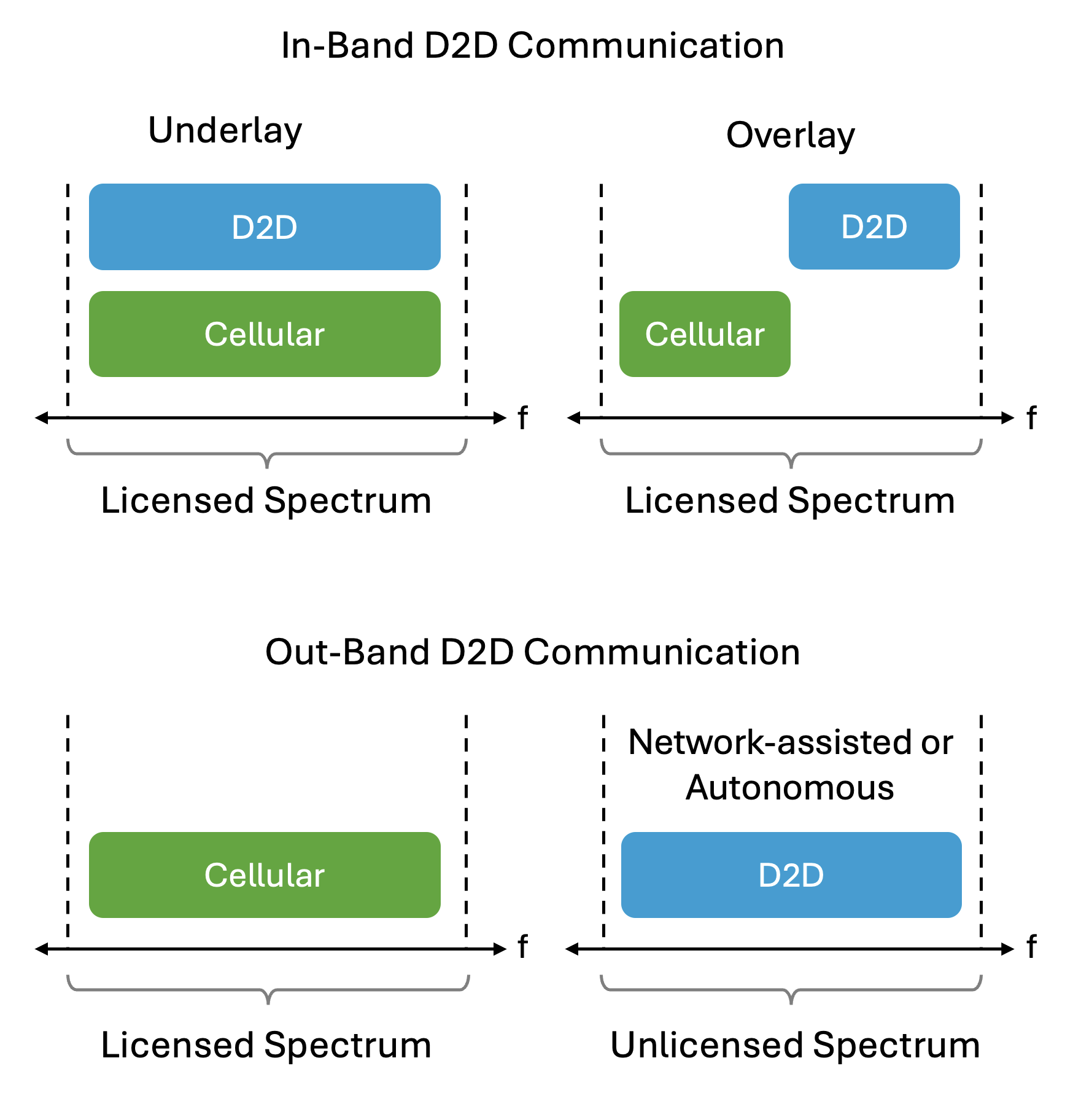}
    \caption{Overlay/underlay in-band D2D vs. controlled/autonomous out-band D2D.}
\label{fig:D2D_IBD_OBD}
\end{figure}

A schematic representation of D2D operation modes in cellular networks is illustrated in Fig.~\ref{fig:D2D_IBD_OBD}. 

\subsubsection{In-Band D2D Communication}
\label{subsubsec:inband_d2d}

In in-band D2D (\gls{IBD}) communication, user devices utilize the licensed spectrum for direct communication while coexisting with traditional cellular users. This approach offers the advantage of tight integration with the cellular infrastructure, allowing network operators to retain control over resource allocation, interference management, and QoS provisioning. IBD is generally divided into two modes: underlay and overlay. 

\begin{itemize}
    \item In the underlay mode, D2D transmissions reuse the same radio resources as cellular users by utilizing techniques such as resource allocation, interference mitigation, and diversity~\cite{Gismalla:2022}. Although this maximizes spectral and power efficiency, it also introduces significant interference challenges, requiring sophisticated interference coordination and power control mechanisms to ensure effective coexistence~\cite{Jameel:2018}.

    \item In contrast, the overlay mode assigns dedicated spectrum resources exclusively to D2D pairs, effectively eliminating interference with cellular transmissions and allowing more signal strength by improved power control especially in relay-assisted networks but at the cost of reduced spectral efficiency~\cite{Gismalla:2022}.  
\end{itemize}

The IBD approach is particularly attractive for proximity-based services where network assistance is essential, such as public safety communications, content sharing, and real-time cooperative applications. However, the performance is heavily influenced by the effectiveness of interference mitigation strategies and the network's ability to manage radio resources dynamically. 

\subsubsection{Out-Band D2D Communication}
\label{subsubsec:outband_d2d}

Out-band D2D (\gls{OBD}) communication enables direct device-to-device transmission over unlicensed spectrum bands, such as those used by Wi-Fi or Bluetooth. By decoupling D2D traffic from the cellular infrastructure, OBD alleviates congestion on licensed bands and enhances the overall spectral efficiency (SE) of the network. Based on the degree of network involvement, OBD can be broadly classified into two categories: controlled and autonomous.
\begin{itemize}
    \item In controlled mode, the cellular network actively coordinates D2D communication by leveraging its advanced management capabilities. It synchronizes D2D users in terms of time, frequency, and phase during cell search procedures and provides essential control-plane functionalities such as device discovery, session initiation, link establishment, resource scheduling, power control, and routing support~\cite{Asadi:2014}. With full visibility into D2D link activity, the network can efficiently perform radio resource management (\gls{RRM}) and mitigate interference between cellular and D2D transmissions. However, these benefits come at the cost of increased signaling overhead and latency due to the need for full channel state information (\gls{CSI}), which becomes particularly burdensome in large-scale deployments~\cite{Liu:2015}.

    \item In contrast, autonomous OBD allows devices to manage their communication independently over unlicensed bands without cellular network assistance. The primary motivation for this approach is to offload traffic from the core network and minimize changes to existing base station infrastructure~\cite{Safdar:2016}. While the network may impose constraints such as maximum allowable transmit power, D2D devices perform peer discovery, link setup, resource scheduling, and power control in a distributed and self-organizing manner~\cite{Asadi:2014}. This mode is applicable in both in-coverage and out-of-coverage scenarios, allowing D2D pairs to communicate without relying on network coordination. Nevertheless, several challenges affect the viability of autonomous OBD, including protocol mismatches between interfaces requiring additional encoding/decoding, elevated security risks due to the unregulated spectrum, and difficulties in guaranteeing QoS. Furthermore, simultaneous D2D and cellular communication demands dual wireless interfaces, which may not be universally supported~\cite{Jameel:2018}.
\end{itemize}

OBD communication is particularly well-suited for proximity-based services such as file sharing, local broadcasting, and multiplayer gaming. However, its performance is influenced by factors such as the availability of unlicensed spectrum, interference from coexisting technologies, communication range, and the EE of the participating devices. 

\begin{figure*}
    \centering
    \includegraphics[width=\linewidth]{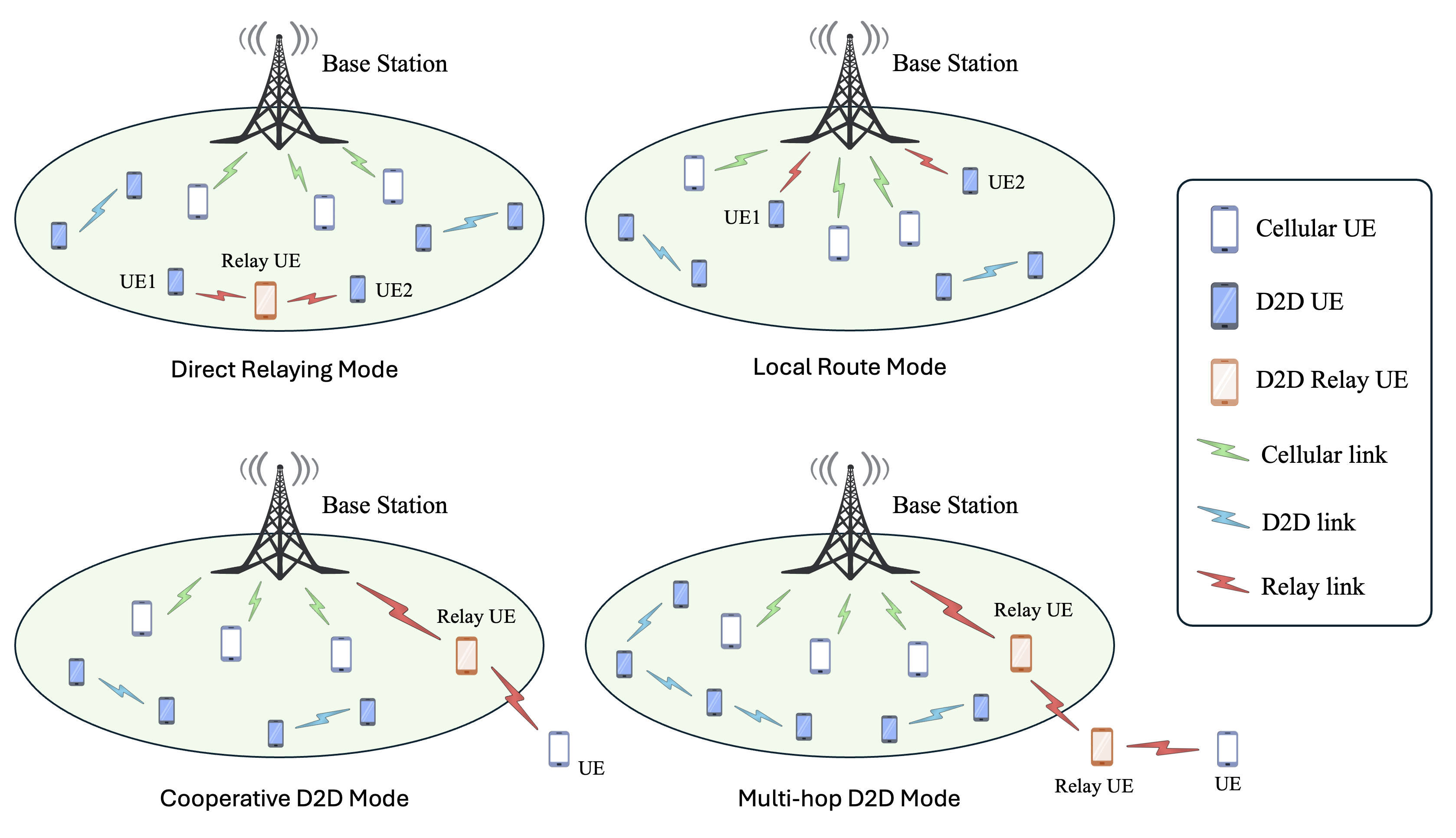}
    \caption{System model of different types of relay mode D2D communication.}
    \label{fig:D2D_relay}
\end{figure*}

\subsection{Relay Mode D2D Communication}
\label{subsec:relay_d2d}

D2D communication faces several challenges when operating within cellular networks, particularly co-channel interference (\gls{CCI}),
which occurs when D2D and cellular transmissions use overlapping spectrum resources. This issue is aggravated when the D2D transmitter and receiver are not in close proximity, resulting in increased outage probability due to poor link quality and limited transmit power, especially in unlicensed spectrum bands~\cite{Amodu:2019}. In such scenarios, conventional interference management techniques like power control often fall short. As a result, relay-based communication has emerged as a promising solution to extend coverage, improve link reliability, and enhance overall network performance. 

To address these limitations, relay-based D2D communication introduces intermediate nodes that assist in forwarding data between communicating devices. These nodes enable multi-hop transmission, thereby improving the feasibility of D2D communication in challenging propagation environments. In general, relay-mode D2D communication can be categorized into two distinct classes: relay-assisted and cooperative D2D communication.

\subsubsection{Relay-Assisted D2D Communication}

In relay-assisted mode, a dedicated or idle UE located between the source and destination acts as a relay to facilitate data transmission. The source node transmits data to the relay, which decodes and forwards the information to the destination node, a technique
commonly referred to as decode-and-forward relaying~\cite{Amodu:2019}. This relaying mechanism helps overcome the path loss and fading experienced in direct links and significantly reduces the outage probability. For efficient relay selection, studies suggest defining an energy-efficient relaying zone, determined by propagation characteristics such as the path loss exponent~\cite{Hourani:2016}. Within this zone, potential relay candidates can be chosen to minimize energy consumption while maintaining link quality. 
Relay-assisted D2D communication includes two practical configurations~\cite{Jameel:2018}:
\begin{itemize}
    \item Direct Relaying Mode: An idle UE directly forwards data between the D2D transmitter and receiver.
    \item Local Route Mode: The BS acts as a relay, forwarding the data from the D2D transmitter to the receiver via uplink and downlink transmissions.
\end{itemize}

\subsubsection{Cooperative D2D Communication}

Cooperative D2D communication takes a complementary perspective by using D2D-enabled devices to support cellular users, particularly those at the cell edge or out-of-coverage area. In this mode, an in-coverage UE assists a cellular user by relaying its data to the BS through a D2D link. This so-called two-hop communication strategy is especially valuable in enhancing the throughput of users experiencing poor channel conditions. Cooperative D2D is part of 3GPP Release 12 and offers notable benefits such as improved signal quality, higher coverage probability, and reduced interference in uplink transmissions~\cite{Amodu:2019}. In addition, cooperative schemes can mitigate the negative impact of D2D interference on cellular users by enabling more intelligent and localized resource reuse~\cite{Lee:2019}.

\subsubsection{Toward Multi-Hop D2D Communication}

While relay-assisted and cooperative modes generally assume a single relay (i.e., two-hop communication), extending this concept to multi-hop D2D offers even greater benefits. In multi-hop scenarios, data is forwarding across multiple intermediate devices before reaching the final destination, enabling coverage extension beyond the direct transmission range. 
Both single-hop and multi-hop relay modes have distinct advantages. Single-hop relaying typically results in lower latency and reduced energy consumption, while multi-hop relaying enhances network connectivity, coverage, and load balancing~\cite{Jameel:2018, Shaikh:2018}. Relay-mode D2D communication thus plays a vital role in realizing the full potential of D2D technology by addressing limitations in range, reliability, and interference management. Fig.~\ref{fig:D2D_relay} illustrates the different relay mode D2D communication scenarios described above.

\subsection{Full-Duplex D2D Communication}
\label{subsec:FD_D2D}

\begin{figure}
    \centering
    \includegraphics[scale=0.25]{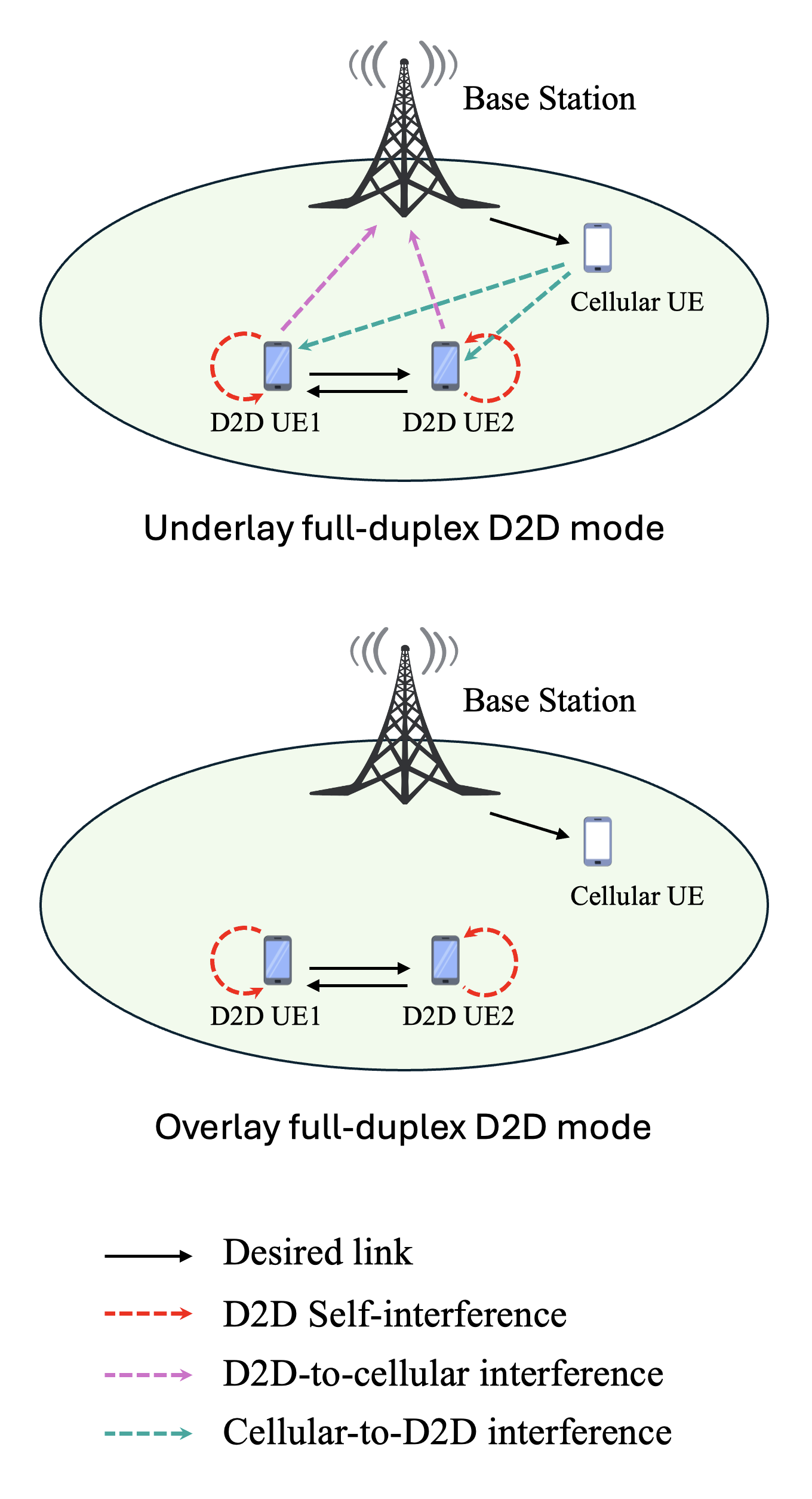}
    \caption{System model of full-duplex D2D communication.}
    \label{fig:D2D_duplex}
\end{figure}

In contrast to traditional half-duplex (\gls{HD}) communication mode, where transmission and reception must occur at different times or on different frequencies, full-duplex (\gls{FD}) communication allows a device to simultaneously transmit and receive signals over the same frequency band. By enabling concurrent bidirectional communication, FD offers the potential to double the SE and reduce latency, making it a promising candidate for next-generation wireless networks, including 5G and beyond.

However, the practical deployment of FD systems introduces significant challenges. Most notably, self-interference (\gls{SI}) occurs when a device's transmitted signal leaks into its own receiver, leading to residual self-interference (\gls{RSI}). In early FD systems, the inability to perfectly cancel RSI posed a serious limitation to achieving reliable FD operation, especially under imperfect CSI conditions. To overcome this, a combination of analog and digital self-interference cancellation (\gls{SIC}) techniques has been developed. These include physical separation of transmit and receive antennas, radio-frequency (\gls{RF}) domain suppression, and baseband digital cancellation~\cite{Liu:2018}. When effectively implemented, these techniques can reduce SI close to the noise floor, enabling more robust FD communication and opening the door to its integration with D2D networks. On the other hand, since D2D users often operate in close proximity and at reduced power levels, SI is inherently less severe, making FD well-suited to boost local throughput and reduce transmission delay. A typical system model of FD enabled D2D in both underlay and overlay modes is depicted in Fig.~\ref{fig:D2D_duplex}.

The integration of FD into cooperative D2D communication can provide additional robustness in complex propagation environments, such as those affected by multipath fading and frequency-selective channels. In such scenarios, FD relaying can enhance system reliability and responsiveness. Nonetheless, mode selection between FD and HD operation becomes a critical challenge, particularly in dense D2D deployments. While FD operation can significantly improve the aggregate data rate of D2D users, it may also lead to increased interference for cellular users. Without proper control, this interference can degrade the QoS for other network participants~\cite{Badri:2021}. 

To manage this trade-off, adaptive mode selection strategies are essential. These strategies dynamically regulate the number of D2D pairs operating in FD mode, balancing throughput gains against the interference imposed on cellular users. Moreover, under light traffic conditions, it may be beneficial for FD capable devices to temporarily revert to HD or even silence mode immediately after completing their transmission tasks, in order to avoid unnecessary interference~\cite{Du:2020}. 

\begin{figure*}
    \centering
    \includegraphics[width=0.8\linewidth]{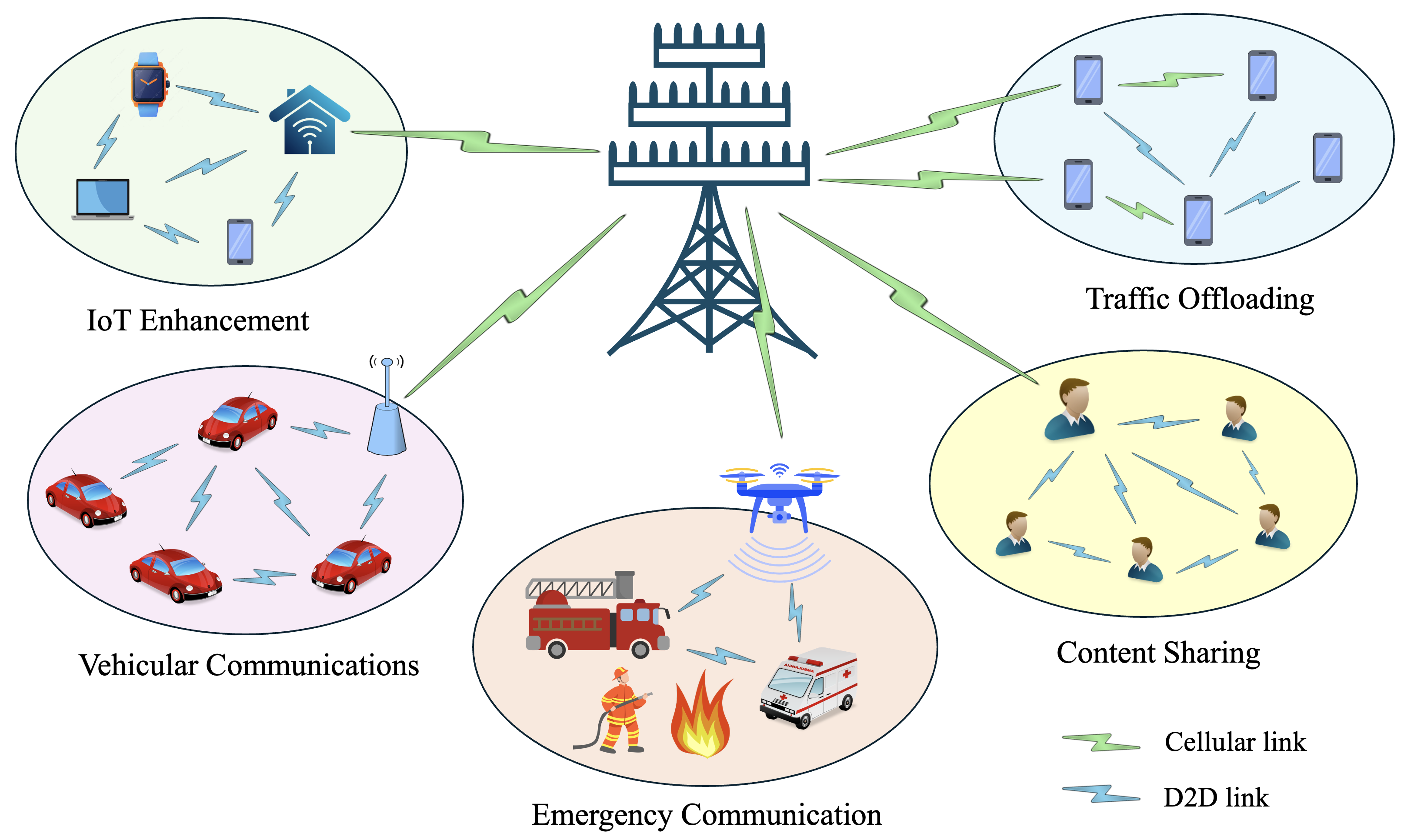}
    \caption{Common applications of D2D communication in 5G and beyond networks.}
    \label{fig:D2D_5G_6G}
\end{figure*}

\subsection{Control Modes of D2D Communications}

Centralized and distributed control architectures represent two fundamental paradigms for managing D2D communication in cellular networks. In a centralized approach, the BS or a central network controller takes charge of key D2D operations such as resource allocation, power control, and mode selection. This method benefits from having a global view of network conditions, enabling optimal coordination between D2D and cellular UEs. Centralized control is particularly effective at mitigating interference and maximizing SE across the network. However, it introduces significant signaling overhead and can become a bottleneck in dense deployments due to limited scalability and increased latency in decision-making processes~\cite{Yin:2016}.

In contrast, distributed control allows D2D UEs to operate more autonomously by making decisions based on local information such as channel state, interference level, and traffic load. This decentralization reduces the need for signaling with the BS and enables more scalable and robust operation, especially in dynamic environments or in scenarios where the BS is unavailable (e.g., out-of-coverage situations). However, distributed control faces challenges related to suboptimal resource use and interference management due to the absence of global coordination. Without a central entity to enforce resource policies, D2D transmissions may overlap and cause significant degradation in network performance~\cite{Ibrahim:2019}.

To combine the benefits of both paradigms, hybrid control mechanisms have been proposed, where initial coordination and policy enforcement are handled by the BS, while real-time operational decisions are made locally by the devices. This hybrid model offers a trade-off between the optimization capabilities of centralized control and the flexibility of distributed methods. By allowing the BS to guide the general framework and letting devices self-adapt to local changes, these architectures are especially promising for next-generation networks where both dense deployments and heterogeneous service requirements must be addressed efficiently~\cite{Maghsudi:2015}.

\section{Interplay of D2D Communication with 5G/6G}
\label{sec:d2d_5G_interplay}

D2D communication has emerged as a key architectural enhancement for both 5G and future 6G networks, offering a paradigm shift from traditional centralized models toward more decentralized and intelligent connectivity. By enabling direct communication between UEs without traversing the core network or BS, D2D significantly reduces latency, enhances SE and EE, and offloads traffic from congested cellular infrastructure. This is particularly beneficial in scenarios such as disaster recovery or network outages, where D2D links can provide resilient, infrastructure-independent communication~\cite{Liu:2024}. 

In 5G, D2D plays a vital role in supporting localized services and autonomous peer-to-peer interaction, with operational modes including direct, relay, and in/out-of-coverage configurations. However, these advantages come with challenges such as interference management, mode selection, spectrum sharing, and peer discovery, especially when operating in underlay mode. Looking ahead, 6G networks hold the promise of being ultra-intelligent, providing a heterogeneous ecosystem that leverages AI-native network management. Within this framework, D2D communication will become even more crucial, benefiting from ultra-dense deployments and the growing computational capabilities of UEs. Future UEs are expected to integrate advanced sensing, computing, and multi-antenna capabilities, enabling them to function as intelligent network nodes\cite{Areqi:2023}. 

Thus, the integration of D2D into 5G and 6G networks not only addresses the increasing demands of emerging applications but also lays the groundwork for autonomous and highly efficient wireless systems. This section explores the advantages of incorporating D2D into 5G/6G architectures, along with the associated technical challenges. Fig.~\ref{fig:D2D_5G_6G} illustrates some key applications of D2D communication in cellular networks.

\subsection{Enhanced Energy and Spectrum Efficiency}

One of the most compelling advantages of device-to-device (D2D) communication is its ability to significantly enhance SE and overall system throughput through localized spectrum reuse. By enabling direct communication between nearby devices, D2D minimizes reliance on centralized BS scheduling and facilitates spatial reuse of cellular resources, allowing multiple concurrent transmissions within the same cell. This localized reuse not only boosts area throughput but also alleviates congestion in the core network which is critical in dense urban environments. Moreover, proximity-based transmissions inherently require less power, which conserves device battery life and reduces energy consumption. Offloading traffic from the cellular infrastructure also lightens the energy load on BSs, especially during periods of peak demand. These energy savings become even more pronounced in scenarios with high user density, where short-range D2D links are far more power-efficient than traditional long-range cellular connections~\cite{Shamaei:2024}.

Advanced D2D strategies, such as relay-based communication, further optimize energy efficiency by selecting routes and relay nodes based on energy-aware criteria. When combined with intelligent power control and energy-efficient scheduling algorithms, D2D-enabled systems can maintain communication quality while operating at minimal power levels. Together, these capabilities make D2D communication a key enabler of green networking goals in 5G and beyond.

\subsection{Reduced Communication Latency}

By enabling direct data exchange between UEs and bypassing the traditional routing through BS or core network infrastructure, D2D plays a vital role in reducing latency within 5G and 6G networks. This local communication path minimizes transmission and queuing delays, leading to faster response times. In time-sensitive applications such as autonomous driving, industrial automation, and augmented reality, this reduction in end-to-end latency is critical for ensuring real-time performance and user safety.

Advanced techniques such as relay-assisted D2D communication and intelligent resource allocation further enhance latency performance. By leveraging MEC in conjunction with D2D, tasks can be offloaded and processed closer to the user, further cutting down on communication and computation delays. Additionally, the growing use of adaptive antenna systems and beamforming in D2D scenarios allows for more efficient and targeted transmissions, reducing retransmissions and improving responsiveness. These capabilities make D2D an essential feature for supporting the ultra-low latency requirements of future wireless systems~\cite{Lin:2019_2}.

\subsection{Support for Emerging Services and IoT Integration}

In the context of the IoT, D2D plays a pivotal role in supporting large-scale, heterogeneous device ecosystem. Many IoT devices operate under strict power and bandwidth constraints. D2D facilitates efficient local information sharing and cluster-based networking among IoT nodes, thereby minimizing signaling overhead and improving scalability. Furthermore, D2D enables collaborative sensing and computing among nearby devices, which is particularly valuable in scenarios such as smart homes, industrial automation, and e-healthcare. It also plays a critical role in massive Machine-Type Communication (mMTC), a core pillar of 5G/6G, by enabling direct and efficient peer-to-peer connections between sensors and actuators in IoT environments~\cite{Giuliano:2024}.

Another advantage of D2D in IoT is its potential to operate in out-of-coverage areas or during infrastructure failure. In such cases, D2D allows IoT devices to form ad-hoc networks and relay data through multi-hop communication, ensuring service continuity and reliability. This is especially critical for mission-critical applications such as disaster response and remote monitoring. Also, context-aware and autonomous D2D interactions will likely be a key ability of future UEs in future 6G networks. This convergence of D2D and IoT technologies will facilitate the realization of intelligent environments with seamless communication across terrestrial, aerial, and satellite domains~\cite{Prasad:2025}.

\subsection{Complementing Advanced Physical Layer Technologies}

D2D communication naturally complements many of the advanced physical layer technologies envisioned for 5G and 6G networks. When integrated with massive MIMO, D2D allows the spatial degrees of freedom (\gls{DoF}s) to be more efficiently utilized by enabling simultaneous transmission without severely impacting the cellular users. Beamforming techniques can further isolate D2D and cellular transmissions, reducing interference and improving link reliability\cite{Amudala:2024}. In mmWave systems, where line-of-sight connectivity is critical and signal attenuation is severe, D2D plays a vital role by offering short-range, high-capacity links that are less susceptible to blockages. Devices in close proximity can establish reliable mmWave D2D connections, enhancing overall coverage and throughput in dense deployments~\cite{Yazdani:2022}. 

D2D also aligns well with RIS technology, which can manipulate the propagation environment to strengthen D2D links, especially in non-line-of-sight conditions~\cite{Fu:2021}. Additionally, integration with NOMA and the physical layer allows D2D pairs to share spectrum more efficiently by leveraging power-domain separation, improving connectivity density and spectrum utilization~\cite{Le:2022}. Altogether, D2D enhances the flexibility and performance of physical layer innovations, making the network more adaptive, efficient, and resilient to environmental and topological changes.

\subsection{Challenges and Design Considerations}

Despite its many advantages, the integration of D2D communication into 5G and 6G networks introduces several critical challenges and risks that must be addressed to fully realize its potential. In D2D underlaid cellular networks where the same spectrum is reused by both cellular and D2D users, the mutual interference can degrade the performance of cellular links and vice versa, making interference management a key design priority. Addressing these issues requires intelligent mechanisms that span mode selection, resource allocation, and adaptive control. For instance, mode selection entails determining whether a D2D link should operate in underlay, overlay, or cooperative relay mode. Resource allocation, meanwhile, involves power control, spectrum assignment, and sub-channel distribution, often requiring these tasks to be solved jointly due to their strong coupling and interdependence~\cite{Cao:2015}.

Various optimization approaches have been proposed for these challenges, including linear and nonlinear programming, game theory, matching theory, graph-based formulations, and an increasing reliance on machine learning, deep learning, and reinforcement learning. The selection of an appropriate control architecture is equally important. Resource allocation and interference mitigation can be performed either in a centralized or distributed fashion. In the centralized mode, the BS, with global CSI, performs holistic optimization to ensure efficient spectrum use and maintain quality of service (\gls{QoS}). 

While this approach is generally effective due to global awareness and coordinated decision-making, it imposes a substantial computational burden on the BS, raising concerns about scalability and robustness in ultra-dense networks. In contrast, the distributed mode delegates control to UEs, allowing D2D devices to locally exchange CSI and make independent decisions. This decentralization improves scalability and fault tolerance but introduces challenges such as decision conflicts, outdated or partial CSI, and potential deviation from system-level objectives. Some contention-based protocols have been proposed to mitigate these issues, but they often suffer from poor SE and weak collision resolution~\cite{WILEY_Jadav:2022}.

A thorough understanding of the foundational technologies shaping 5G and emerging 6G networks is essential to grasp how D2D communication can be effectively integrated to enhance overall network performance. While this integration unlocks a wide range of opportunities, it also brings significant design and operational challenges. To investigate these aspects in detail, the following sections are structured around the interplay between D2D communication and key enabling technologies for 5G/6G. Each section will begin with a brief overview of the respective technology and then analyze how D2D complements it, the main technical challenges involved, and relevant insights drawn from recent research efforts aimed at overcoming these challenges.

\section{Interplay between D2D and Massive MIMO}
\label{sec:d2d_massive_mimo}

\begin{figure*}
    \centering
	\includegraphics[width=\linewidth]{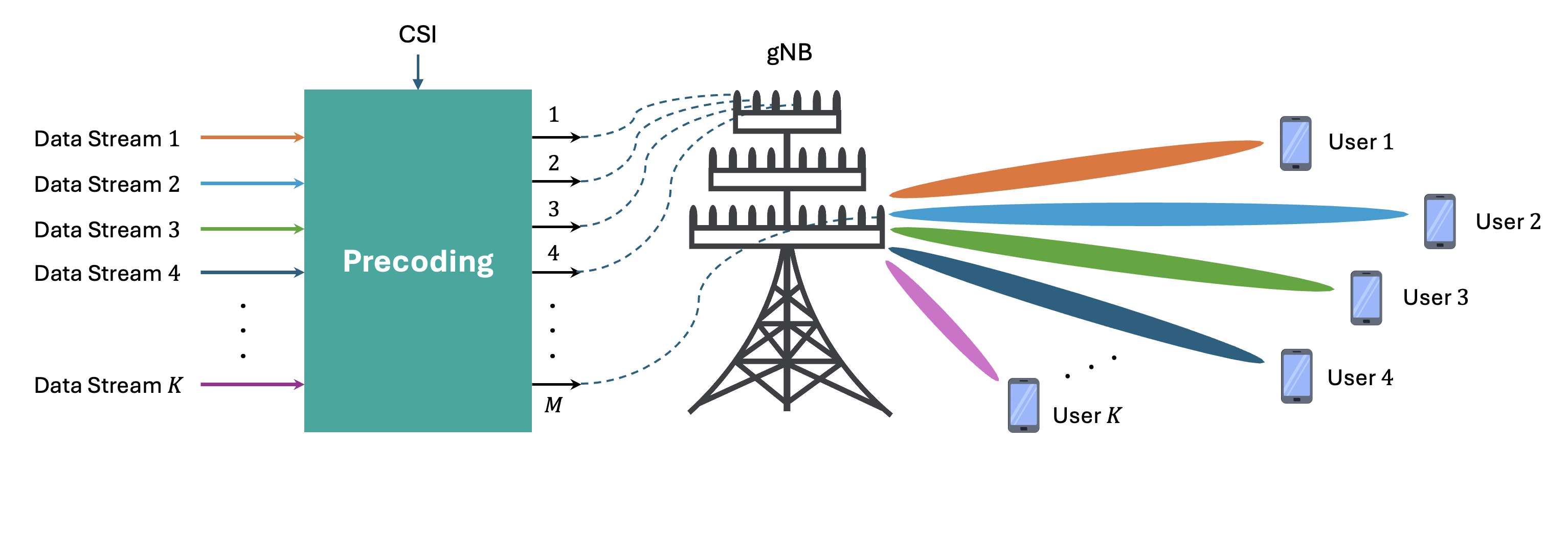}
	\caption{Precoding in a massive MIMO system with $M$ antennas at BS communicating with $K$ users.}
	\label{fig:massive_mimo}
\end{figure*}

\subsection{Overview of Massive MIMO}

MIMO technology has become a foundational component of modern wireless communication systems and is widely adopted in standards such as IEEE 802.11 ac/ax (Wi-Fi), WiMAX (IEEE 802.16e), LTE/LTE-Advanced, and more recently, 5G NR. The concept of MIMO initially emerged from point-to-point configurations, where both the transmitter and receiver are equipped with multiple antennas to enhance data rates through spatial multiplexing. This evolved into multi-user MIMO (\gls{MU-MIMO}), where a BS with multiple antennas simultaneously serves several single-antenna users, thereby improving SE without requiring complex hardware at the user end~\cite{Marzetta:2015}. 

While MU-MIMO provides notable benefits in terms of cost-efficiency and reduced dependency on propagation conditions, it introduces inter-user interference, which can significantly degrade system performance. Addressing this challenge typically involves complex signal processing techniques for user scheduling, channel estimation, and interference suppression~\cite{Fatema:2018}. However, with the advent of massive MIMO, where the number of antennas at the BS significantly exceeds the number of served users, many of these issues can be mitigated. In massive MIMO systems, the channels between the BS and different users become asymptotically orthogonal, allowing simple precoding schemes such as matched filtering or zero-forcing to achieve near-optimal performance~\cite{Lin:2015}.

Massive MIMO not only enhances the capacity and SE of the network but also improves EE by enabling focused beamforming and reducing the need for dense BS deployments. The large spatial DoFs available in such systems support fine-grained user separation, robust interference management, and improved performance for edge users who typically suffer from low signal-to-noise ratios (\gls{SNR}s)~\cite{Marzetta:2015}. These capabilities make massive MIMO a critical enabler of the high-capacity, energy-efficient, and interference-resilient design goals of 5G and emerging 6G networks. Fig.~\ref{fig:massive_mimo} illustrates precoding in a massive MIMO system. 

\subsection{Integration of D2D with Massive MIMO}

As 5G and beyond aim to support ultra-dense, high-capacity networks with low latency, integrating D2D communication with massive MIMO emerges as a promising solution for achieving these goals. This integration is particularly valuable in scenarios like vehicular networks, where D2D offers low-latency communication, and massive MIMO ensures reliable coverage and capacity.

D2D communication allows nearby devices to communicate directly, without going through the BS or core network. This direct link reduces transmission delay, improves SE, and saves energy, especially because D2D pairs can use lower power and higher data rates thanks to their short distance. Offloading some of the cellular traffic to D2D links can also reduce network congestion and enhance overall system performance. Combining D2D with massive MIMO, where BSs are equipped with large antenna arrays, can be a powerful approach. Massive MIMO enables highly directional transmissions, creating near-orthogonal channels that help minimize interference. This has led to the belief that D2D communications in massive MIMO systems could be straightforward and efficient~\cite{Lin:2015}. 

However, interference is still a major challenge, especially in underlay D2D mode where D2D pairs reuse the same frequency resources as cellular users. Even though the BS can use its antenna array to suppress interference from D2D users, D2D receivers still suffer from interference caused by cellular users, especially when many users are active at the same time. As the number of active cellular users increases, D2D links sharing the same spectrum face more interference. To protect D2D transmissions, one may need to limit the number of active cellular users, but doing so reduces the potential gain of massive MIMO. This creates a trade-off between supporting D2D and maintaining high capacity in the cellular network~\cite{Nie:2021}.

Another challenge is the high density of D2D users, which is often needed to offload enough traffic. But a large number of D2D links can cause significant interference to cellular users, degrading network performance. One possible solution is to use D2D devices with multiple antennas so they can perform beamforming to reduce interference. However, this is not always practical. Most D2D devices, like smartphones and IoT gadgets, are limited in size, power, and processing capability, making it hard to support beamforming. Even if more antennas improve D2D link quality, they do not necessarily help reduce interference at the BS~\cite{Senadhira:2016, He:2017, Agarwal:2018, Song:2021}. Moreover, the presence of dynamic D2D links also complicates channel estimation in massive MIMO systems, possibly increasing pilot contamination and CSI overhead, especially when D2D pairs occur more frequently. Still, massive MIMO can handle interference from D2D links fairly well, and D2D transmissions are often robust against interference from cellular users, even when they share the same channel~\cite{Shalmashi:2016}. 

In summary, while D2D and massive MIMO can complement each other and boost network performance, their integration requires careful interference management, especially when many D2D users are active. Without this coordination, the benefits of massive MIMO can be compromised. 

\subsection{Mode Selection}

Mode selection is a fundamental step in integrating D2D communication into cellular networks, especially in the presence of massive MIMO systems. It involves determining the operational mode for each D2D pair (underlay, overlay, relay-assisted, or cellular mode) based on current channel conditions, interference levels, and traffic demands. Effective mode selection improves the SE and EE of both D2D and cellular users while minimizing interference and ensuring QoS requirements are met~\cite{Cao:2015}. 

In underlay in-band D2D, the D2D pair shares the spectrum with cellular users, often reusing the uplink resources. This improves spectrum utilization but introduces interference, which is especially critical in massive MIMO systems where cellular users are served simultaneously. In contrast, overlay in-band D2D assigns dedicated resources to D2D pairs, avoiding interference but at the cost of lower spectrum efficiency. The relay mode uses D2D users to assist communication for others or extend coverage, while the cellular mode routes traffic entirely through the BS, treating the D2D pair as regular cellular users~\cite{Lin:2019}. 

The choice of mode must balance the interference trade-offs and depends heavily on the network load. For instance, when traffic is light, idle cellular resources can be used by D2D links without causing harmful interference. However, in high-traffic conditions, not all D2D pairs can receive dedicated resources, necessitating efficient reuse strategies through underlay modes. Therefore, a mode selection algorithm should consider the quality of both D2D and cellular links, current interference conditions, and user requirements~\cite{Jayasinghe:2014}. 

Because mode selection interacts closely with spectrum allocation and power control, these tasks are often solved jointly. A prerequisite for effective decision-making is accurate CSI for both cellular and D2D links at the beginning of each frame. This requirement becomes more stringent in massive MIMO systems due to the higher number of simultaneously served users and the increased sensitivity to interference patterns~\cite{Zhou:2023}. 

Existing mode selection strategies can be broadly classified based on network size and execution architecture (centralized vs. distributed). In small-scale networks, exhaustive search combined with optimal power allocation may yield high performance. However, such an approach becomes impractical as the number of users increases. In larger networks, cross-layer optimization frameworks have been developed, incorporating mode selection alongside resource allocation and interference coordination. To handle the scalability issues, researchers have proposed a variety of algorithmic approaches, including game theory and graph theory for distributed and structured decision-making, respectively, nonlinear optimization for system-wide performance goals, and ML/RL for adaptive and data-driven solutions.

Most traditional studies treat cellular users as primary and D2D users as secondary, echoing the principles of cognitive radio. These models prioritize cellular users' QoS and limit D2D spectrum reuse, usually to the uplink. However, recent research has begun exploring more balanced models where D2D and cellular users are treated equally in mode selection and resource sharing, enhancing fairness and network flexibility. Ultimately, mode selection plays a pivotal role in enabling the coexistence of D2D communication and massive MIMO systems. Its effectiveness directly influences network throughput, user experience, and the feasibility of D2D under dense deployment conditions.

A mode selection solution for D2D-enabled cellular networks was proposed by~\cite{Klugel:2020}, which utilizes the Generalized Benders Decomposition technique to decouple the optimization problem of PHY and MAC layers, while the mode selection decisions are made in a central coordinator node. A D2D mode selection scheme based on the Fuzzy clustering algorithm was proposed by~\cite{Algedir:2020}, in which the D2D users are switched between dedicated and/or reuse mode based on the state of the RB availability. The authors of~\cite{Han:2020} investigated the mode selection problem for FD-enabled D2D communications and proposed solutions for the optimization of FD/HD overlay/underlay operation modes. A stochastic geometry approach for mode selection and interference management of FD-enabled D2D communications was proposed by~\cite{Badri:2021}. An energy-aware mode selection mechanism for D2D-enabled 5G network was investigated by~\cite{Tsai:2023}. It divides users into three tiers based on their distance from the BS, and allocates the resources to different tiers using the Hungarian algorithm. A hybrid scheme for D2D-enabled 5G heterogeneous networks (\gls{HetNet}s) was proposed by~\cite{Feng:2020}. It employs online ML for throughput maximization and RL for mode selection. D2D communications reuse the uplink resources by controlling the interference. In~\cite{Du:2020}, a duplex mode selection for D2D-aided underlaying cellular network
was proposed in which the D2D users can switch between HD and FD operation mode based on their channel quality. Reference~\cite{Vu:2021} investigated the integration of full-duplex D2D communication into a massive MIMO system and utilized joint beamforming and power allocation design to alleviate the interference between cellular and D2D transmissions and to maximize the average sum-rate.

\subsection{Spectrum Allocation and User Scheduling}

In D2D-enabled massive MIMO systems, spectrum allocation and user scheduling are two tightly coupled processes essential for managing interference, enhancing spectral and energy efficiency, and ensuring QoS for both cellular and D2D users. Their joint optimization becomes increasingly important due to the large number of antennas at the BS and the dense deployment of users in 5G and beyond networks~\cite{Tehrani:2014}.

Spectrum allocation determines how frequency resources are distributed among cellular and D2D users. In underlay D2D scenarios, D2D pairs reuse the cellular spectrum, which improves spectrum utilization but introduces mutual interference. The challenge is to allocate spectrum in a way that maximizes overall network performance while controlling interference. User scheduling, on the other hand, determines which users (cellular or D2D) are allowed to transmit in a given time slot and frequency band. In massive MIMO systems, the spatial degrees of freedom can be exploited to serve many users simultaneously via beamforming. However, interference among co-scheduled users, especially between D2D and cellular users sharing the same resources, must be carefully controlled~\cite{WILEY_Ali:2017}.
 
Spectrum allocation is typically jointly optimized with user scheduling, as these two processes are tightly coupled. This joint optimization can be executed in either centralized or distributed fashion. In centralized schemes, the BS collects CSI and controls spectrum assignment across both D2D and BS-to-device (B2D) links. These schemes often achieve better performance in terms of SE and EE due to their global network view and coordination capabilities. In distributed schemes, on the other hand, UEs perform local decision-making for spectrum reuse based on limited CSI or contextual information. While more scalable and suitable for large or dynamic networks, distributed approaches typically suffer from suboptimal performance due to lack of coordination and global awareness~\cite{Du:2021}. 

The authors of~\cite{Du:2021} used a deep Q-network (\gls{DQN}) scheme to jointly allocate resources to the D2D users and control their power according to their mode of operation. A joint mode selection and resource allocation algorithm for D2D users was investigated in~\cite{Yan:2019} to maximize the system throughput while considering the energy harvesting and QoS constraints of cellular users. In~\cite{Jeon:2021}, a mode selection and resource allocation algorithm for FD-enabled D2D communication systems is proposed that performs the selection of D2D pairs based on concepts from graph theory. A rate adaptation method based on a stochastic geometry approach is presented in~\cite{Chen:2017}, which minimizes the interference from D2D to cellular users in the multi-cell scenario. An interference-aware spectrum sharing strategy that jointly coordinates the scheduling decisions, beamforming, and power control among both the cellular and D2D links is adopted in~\cite{Shen:2019}, wherein the weighted sum-rate maximization problem is solved using matrix fractional programming. A learning-based, robust rate scheduling framework that maximizes the SE of a D2D underlaid massive MIMO system is proposed in~\cite{Zhang:2019}, whereby both non-coordinated and coordinated channel estimation schemes are analyzed. Aiming to design an online controller that dynamically schedules users and configure their links to minimize the delay, an RL-based joint user-scheduling, relay-selection, codebook optimization, and beam tracking algorithm is proposed by~\cite{Zhang:2023}.

\subsection{Power Allocation}

\begin{figure}
 \centering
	\includegraphics[width=0.7\linewidth]{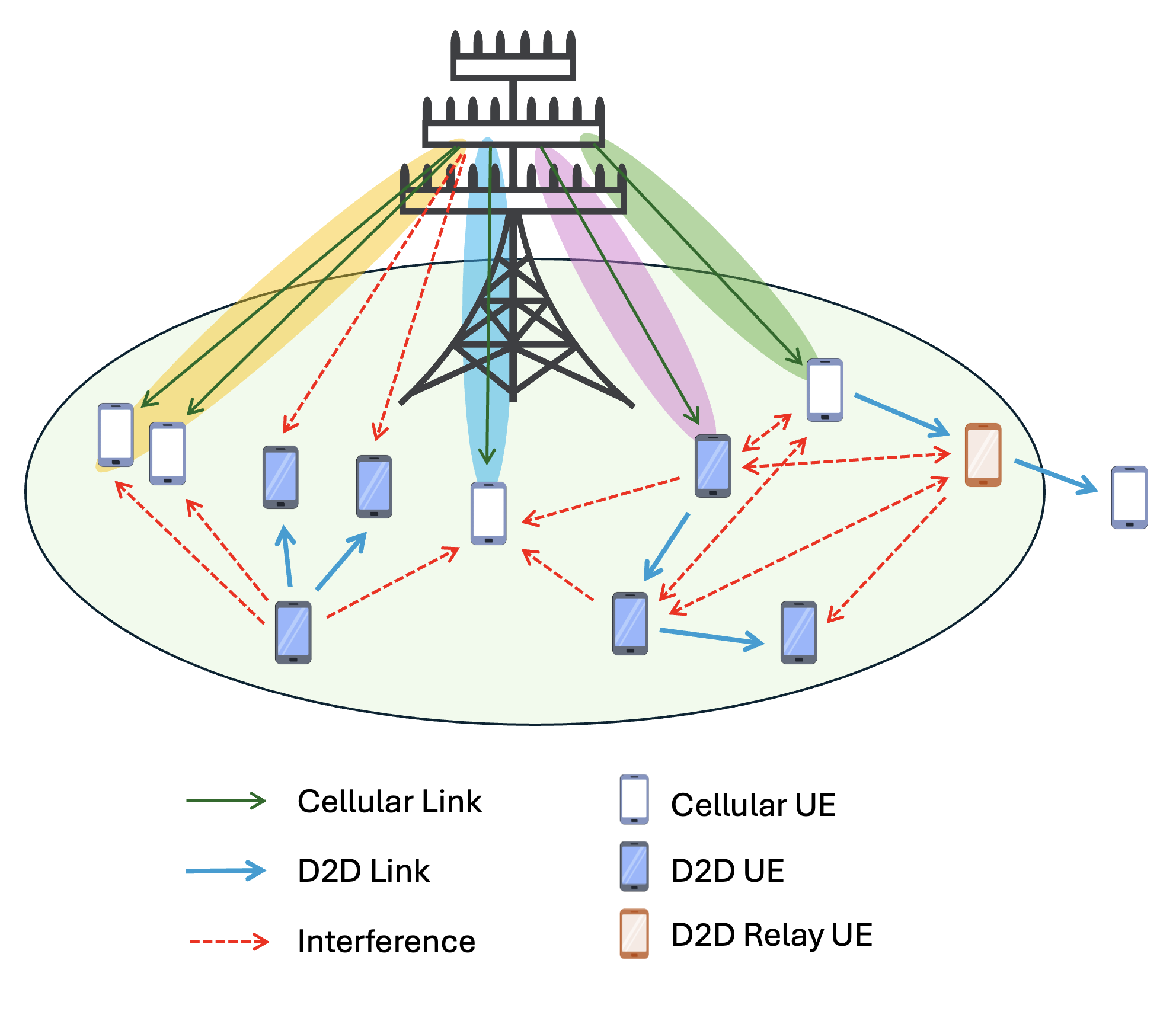}
	\caption{D2D underlay in a massive MIMO-enabled cellular network.}
	\label{fig:d2d_massive_mimo}
\end{figure}

Power allocation is a fundamental technique used to mitigate interference in D2D-enabled cellular networks, particularly when D2D links operate as an underlay to the cellular infrastructure. Without proper power control, D2D transmissions can cause severe co-channel interference (\gls{CCI}) to cellular users, especially due to the near-far problem, where strong signals from nearby D2D users overshadow weaker signals intended for distant users. To ensure the QoS for cellular users, effective power allocation strategies are required that regulate the transmission power of D2D users based on network conditions and interference constraints. Power allocation is also performed in centralized and distributed modes~\cite{Ghazanfari:2019}. A massive MIMO enabled cellular network with underlaid D2D transceivers is illustrated in Fig.~\ref{fig:d2d_massive_mimo}.

In centralized schemes, the BS plays a key role by first assigning radio resources and then adjusting the transmission power of D2D users based on the available CSI. By leveraging its global network view and computational capabilities, the BS can optimize D2D power levels to minimize interference while maximizing overall spectral efficiency. Centralized power control prefers uplink resource reuse for D2D communication, as uplink resources are often underutilized compared to downlink resources, and the BS is better equipped to manage interference originating from D2D transmitters. This approach also allows the BS to dynamically adapt power settings based on real-time CSI, ensuring minimal degradation in cellular users' QoS~\cite{Memmi:2017}.

In contrast, distributed power allocation allows D2D users to adjust their transmission power autonomously based on local information, such as direct channel gains and predefined thresholds. These schemes aim to minimize individual power consumption while preserving link quality and reducing aggregate interference. Although distributed solutions are more scalable and require less overhead than centralized ones, their performance may be suboptimal due to the lack of global network knowledge~\cite{WILEY_Ali:2017}.

Power control strategies are highly dependent on the spatial distribution of users and the availability of accurate CSI. Traditional models often represent user locations using Poisson Point Processes (\gls{PPP}s), assuming a uniform spatial distribution. However, in practical D2D scenarios, users tend to form clusters around hotspots, urban centers, etc. To reflect this, recent studies have adopted Poisson Cluster Process (\gls{PCP}) models, providing a more realistic view of D2D link behavior and interference patterns~\cite{Turgut:2019}.

One of the major challenges in both centralized and distributed power control is the acquisition of accurate and timely CSI. Channel estimation errors, feedback delays, and user mobility make it difficult to obtain perfect CSI, especially for links between D2D pairs. Imperfect CSI can lead to inefficient power allocation, increased interference, and reduced system performance~\cite{Memmi:2017}.

The authors of~\cite{Qiao:2021} proposed a power control scheme for both D2D and cellular users in order to maximize the average sum-rate of a D2D-enabled cell-free downlink massive MIMO system when only imperfect CSI is available and access points (\gls{AP}) utilize low-resolution digital-to-analog converters (\gls{DAC}s). A hybrid framework proposed in~\cite{Librino:2020} performs the power allocation for D2D devices at the beginning of each time slot and in distributed manner, while the joint channel and model selection for D2D devices is done in centralized mode 
at a much longer time scale to reduce the signaling overhead. A joint mode selection and power control by considering the interference region for interference management of D2D-enabled heterogeneous cellular network was
studied by~\cite{Liu_2:2020}. Employing the deep deterministic policy gradient (\gls{DDPG}) approach, \cite{Zhang:2021} proposed a joint mode selection, resource allocation, and power control for D2D-enabled HetNet with the goal of EE maximization. A power control scheme at both cellular and D2D users with fixed beamformers was proposed in~\cite{Ghazanfari:2019} considering two different objectives: maximizing the minimum SE and minimizing the product of SINRs. 

\subsection{CSI Acquisition and Channel Estimation}

Accurate CSI is fundamental to the performance of massive MIMO systems, particularly for effective beamforming, interference management, and user scheduling. CSI acquisition becomes even more challenging when massive MIMO is integrated with underlay D2D communication due to the increased user density and more complex
interference dynamics. CSI acquisition in massive MIMO is highly dependent on the duplexing mode~\cite{Sheikhi:2020}. Massive MIMO systems operate in either time division duplex (\gls{TDD}) or frequency division duplex (\gls{FDD}) modes. TDD systems are generally preferred because they can exploit channel reciprocity, allowing BSs to estimate downlink CSI from uplink pilots, which simplifies the channel estimation process significantly. In contrast, FDD systems, which still dominate commercial cellular deployments due to spectrum licensing policies, suffer from high overhead in CSI feedback as downlink CSI must be explicitly estimated and fed back through uplink control channels~\cite{Nie:2021}.

In TDD systems, orthogonal uplink pilot signals are typically used for CSI estimation. However, in multiuser scenarios, limited pilot resources lead to pilot reuse, which causes pilot contamination, a key limiting factor in massive MIMO performance. This issue worsens when D2D communication coexists with massive MIMO systems because D2D users also require pilot resources for their channel estimation, further increasing the overhead. Since D2D links are typically short-range and randomly distributed within a cell, assigning the same pilot to spatially distant D2D links has been considered a feasible compromise to reduce overhead while maintaining tolerable interference levels~\cite{Xu:2018}. 

To reduce overhead, pilot reuse strategies have been proposed in which orthogonal pilots are reused by users sufficiently separated either within the same cell or across neighboring cells. Given the low transmission power and short range of D2D links, assigning identical pilot sequences to spatially distant D2D pairs can cause tolerable levels of contamination. Nonetheless, D2D transmissions are unpredictable and may still disrupt cellular CSI acquisition. To manage this, open-loop power control with maximum transmission power constraints is commonly employed. This approach aims to regulate the average interference caused by D2D transmissions by ensuring that the received D2D signal power remains within acceptable bounds. However, such control mechanisms require some degree of global CSI at the BS, which is not always feasible~\cite{He:2017}.

Co-channel D2D links also degrade CSI estimation accuracy, further impacting resource allocation and scheduling decisions. A simple mitigation strategy involves deactivating D2D transmissions during the cellular training phase, though this reduces the SE of D2D links. Consequently, various pilot scheduling techniques have been proposed to alleviate pilot contamination, such as a time-shifted pilot allocation scheme~\cite{Fernandes:2013}, a smart allocation scheme based on max-min fairness problem~\cite{Zhu:2015}, a deep learning-based scheme~\cite{Kim:2015}, a graph theory-based solution~\cite{Zhu_2:2015}, area-division schemes~\cite{Gao:2018}, and a swarm-intelligence based optimization algorithm~\cite{Nie:2021}.

For FDD-based massive MIMO systems, the main barrier lies in the high feedback overhead associated with downlink CSI acquisition. Since the BS cannot infer downlink channels from uplink pilots, users must estimate the downlink channel and feed it back through the uplink. However, as the number of antennas increases, the size of the channel vector grows proportionally, making full-resolution feedback impractical. In practice, predefined codebooks are used to quantize channel vectors, which are then fed back subject to limited uplink budgets. A finer quantization improves precoding accuracy but consumes more feedback resources, while coarser quantization degrades performance due to increased downlink interference~\cite{Qiu:2019}. 
Various schemes have been proposed to reduce the feedback overhead of FDD-based massive MIMO systems, such as advanced trellis-extended codebook design~\cite{Choi:2015}, compressive sensing (\gls{CS}) based approaches~\cite{Gao:2015}, antenna grouping-based feedback reduction techniques~\cite{Lee:2015}, downlink channel reconstruction via frequency-independent parameters of multi-path delays~\cite{Han:2018}, angular-domain energy distribution-based channel estimation~\cite{Zhang:2018}, rate splitting encoding at the massive MIMO transmitter~\cite{Dai:2016}, a two-layer precoding structure exploiting the low-rank property of the channel covariance matrices~\cite{Kim:2015}, cluster-based two-stage beamforming~\cite{Song:2022}, etc.

One notable class of techniques uses channel covariance matrices to reduce instantaneous feedback requirements. In particular, users with similar channel covariance structures (often due to similar locations or surrounding environments) can be grouped into clusters. Each cluster can then be served via statistical beamforming by projecting signals onto orthogonal subspaces associated with each group. This reduces both feedback and interference~\cite{Adhikary:2013}. A cluster-based two-stage beamforming framework allows the BS to use the same inner precoder for users with identical covariance matrices, while only the effective channels are used to design the outer precoder~\cite{Qiu:2019}. To refine this further, covariance-based hybrid feedback schemes have been proposed, which assign different amounts of feedback bits to users based on pairwise covariance orthogonality~\cite{Cottatellucci:2016}.

Despite these advancements, even covariance-based approaches can incur considerable feedback overhead when real-time adaptation is required. Therefore, more recent efforts aim to exploit synergies between D2D communication and massive MIMO. In dense networks, D2D links can assist not only in data delivery but also in signaling and CSI exchange. For example, in a precoder feedback scheme, users exchange their local CSI via high-rate D2D links to reconstruct global CSI and collaboratively compute the precoder, which is then fed back to the BS through a rate-limited uplink channel. While this reduces the uplink burden, it relies on perfect and instantaneous D2D links, which are impractical in most real-world systems due to quantization noise and delay~\cite{Chen:2017}. 

\begin{figure}
   \centering
	\includegraphics[width=0.7\linewidth]{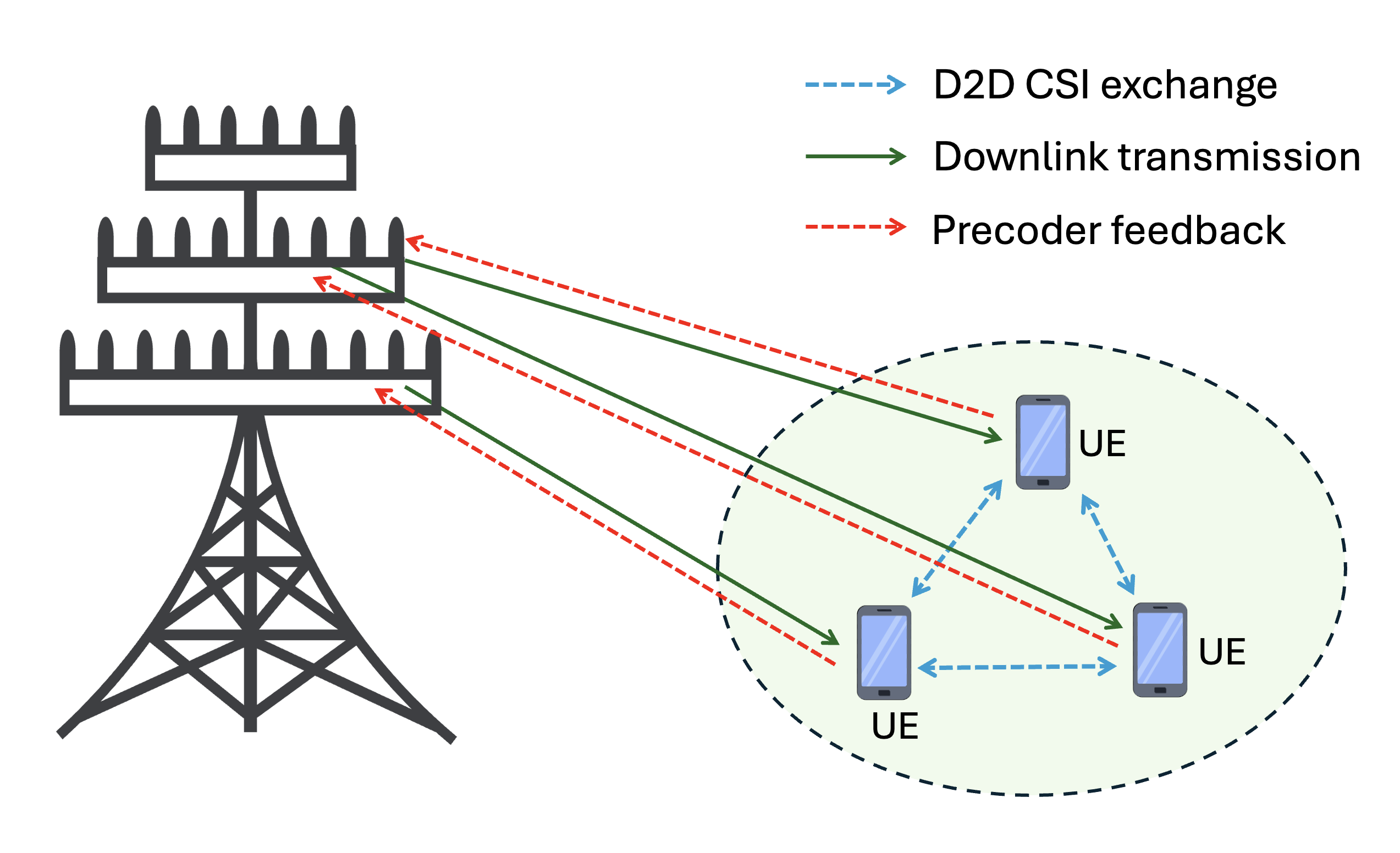}
	\caption{D2D cooperation within a cluster for CSI exchange and precoder feedback to the BS.}
	\label{fig:d2d_precode_feedback}
\end{figure}

Another promising solution is cascade precoding, where the precoding matrix is decomposed into inner and outer components. Fig.~\ref{fig:d2d_precode_feedback} illustrates the idea of cascade precoding by aid of D2D links. The inner precoder is designed based on the eigenstructure of the downlink spatial covariance matrix, which is often shared among users in the same cluster due to similar propagation environments~\cite{Qiu:2019}. This transforms the high-dimensional physical channel into a low-dimensional effective channel on which traditional CSI estimation and feedback schemes can operate more efficiently. Furthermore, users sharing the same inner precoder can establish high-quality D2D links to exchange receiver-related information, enhancing cooperative signal processing~\cite{Cottatellucci:2016}. 

A cooperative precoder feedback scheme was proposed by~\cite{Chen:2017} in which D2D links only exchange a portion of CSI that lies in the overlapping signal subspace. This scheme outperforms the CSI feedback schemes even with limited D2D capacity and moderate delays for CSI exchange. In~\cite{Chen_2:2017}, a dual regularization scheme 
was proposed to unify the CSI and precoder feedback schemes in which D2D links compute the feedback vector to minimize the expected mean squared error (\gls{MSE}) based on their link quality. The BS also regularizes the exchanged feedback from D2D users to minimize the MSE. The proposed scheme is more robust against channel uncertainties and achieves a higher sum-rate than both individual schemes while not affecting non-D2D users. A D2D-relay-assisted scheme was proposed by~\cite{Cottatellucci:2016} in which users are grouped in different clusters based on the similarity of their effective channel subspace. Data streams for a target user in a cluster are transmitted through the beamforming and then all non-target users in that cluster amplify and forward that received signal, thus performing as a virtual MIMO system where the target user will receive multiple independent versions of the transmitted signal. In~\cite{Liu:2017}, a decoding codebook-based approach for D2D-enabled cooperation was proposed that can enhance achievable performance without the need for CSI at the BS. A cross-layer optimization
algorithm that maximizes the total utility function with aid of D2D users' cooperation was presented in~\cite{Liu:2020}. 


\subsection{Interference Alignment}

Interference is one of the most critical challenges in multi-tier and densely populated networks, especially in massive MIMO systems where many users are served simultaneously using spatial multiplexing. While massive MIMO inherently provides substantial spatial DoFs to suppress interference through beamforming, the complexity of interference grows significantly in heterogeneous and dense deployments, particularly due to interference among users in the same tier as well as the interference between macro and small cells. 

Interference alignment (\gls{IA}) is a well-established technique that addresses this challenge by aligning the interference signals from different transmitters into a lower-dimensional subspace at each receiver. This leaves the remaining subspace interference-free for decoding the desired signals. In massive MIMO systems, the high dimensionality offered by large antenna arrays makes IA more feasible, as it becomes easier to find subspaces where interference can be confined without sacrificing the signal space needed for desired transmissions~\cite{Krishnamurthy:2015}.

However, practical implementation of IA in massive MIMO networks is challenging. A key limitation is the requirement for full and precise global CSI at both transmitters and receivers. This is especially problematic in massive MIMO, where the number of antennas and users is high, leading to excessive CSI feedback overhead and synchronization complexity. Moreover, in heterogeneous networks, the asymmetry in transmission power, antenna configuration, and spatial distribution among macro cells, small cells, and user devices further complicates IA feasibility and design.

This is where D2D communication emerges as a valuable enabler for practical IA. D2D links allow user devices to directly exchange partial CSI or coordinate their transmission strategies without routing signaling through the BS. For example, in a macro cell environment with embedded small cells, users can form local D2D clusters to cooperatively design their precoders and receive filters. By interacting directly, users can estimate effective interference channels and align their interference locally, reducing the burden on the base station and minimizing global CSI requirements. Moreover, D2D support enables the system to adapt to heterogeneous connectivity and channel conditions. For instance, in scenarios where some users have limited antenna resources or weak links to the base station, D2D communication allows them to leverage better-connected peers to relay CSI information or participate in joint interference management. This distributed cooperation significantly enhances the robustness of IA schemes in practical deployments~\cite{Zeng:2018}.

The effects of spatial dependencies in MIMO interference networks, which appear as rank-deficient channel matrices, was studied in~\cite{Krishnamurthy:2015}. The authors of~\cite{Jiang:2013} proposed using IA to improve EE in a MIMO downlink network with underlaying D2D communication. In~\cite{Zeng:2018},
a D2D-assisted IA framework was proposed that reduces the CSI overhead by enabling direct CSI exchange and coordination among user devices in a massive MIMO system. By leveraging D2D links, a user grouping strategy was introduced that enables joint design of the precoder and receive filters to improve interference management scalability as user density increases.

\begin{figure}
   \centering
	\includegraphics[width=0.7\linewidth]{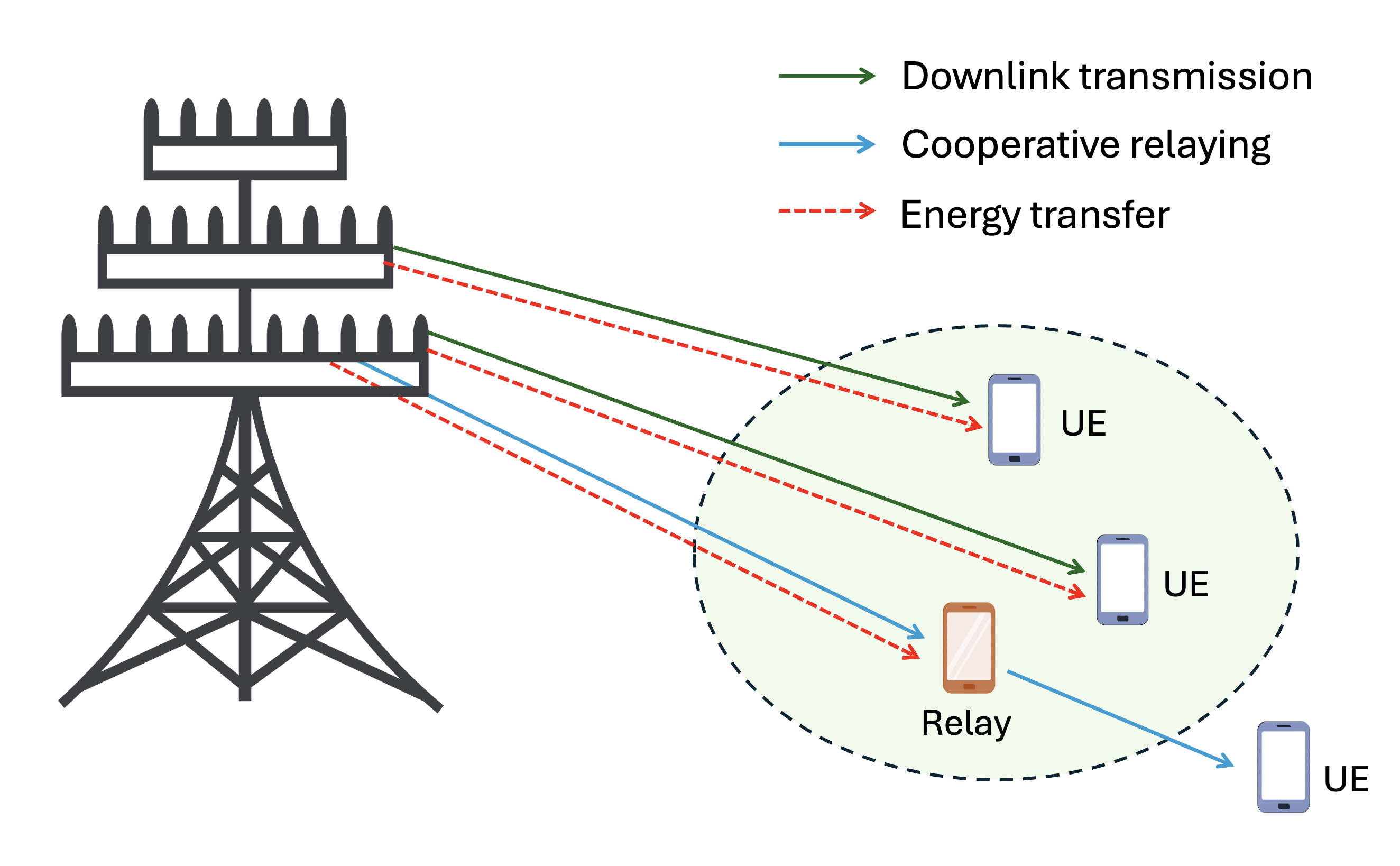}
	\caption{SWIPT with cooperative relaying for a massive MIMO system.}
	\label{fig:d2d_swipt}
\end{figure}

\subsection{Energy Harvesting}

The integration of energy harvesting (\gls{EH}) into D2D communication within massive MIMO networks offers a promising direction for enabling self-sustaining and energy-efficient communication systems, particularly in Beyond 5G scenarios. As the scale of wireless devices continues to grow, especially with the proliferation of D2D and IoT deployments, the energy demands of the network significantly increase. Many of these devices, especially those deployed in remote or inaccessible environments, are battery-powered and face challenges related to frequent recharging or replacement. Energy harvesting provides an attractive solution by allowing these devices to scavenge energy from ambient sources~\cite{Pradhan:2021}.

Massive MIMO systems, with their large antenna arrays and focused beamforming capabilities, can serve as reliable and controllable sources of RF energy for harvesting. This is particularly beneficial for enabling Simultaneous Wireless Information and Power Transfer (\gls{SWIPT}) in D2D-enabled networks, where user devices can decode information and harvest energy from the same RF signal (Fig.~\ref{fig:d2d_swipt}). Such an approach eliminates the need for separate power transfer mechanisms and enhances SE. The energy harvested from massive MIMO base stations or D2D transmitters can be used to power low-energy D2D devices, enabling them to operate autonomously for extended periods. Moreover, the dense deployment of massive MIMO infrastructure in urban areas ensures the availability of strong RF signals, making RF-based EH a practical and scalable solution. SWIPT-based relaying schemes also gain importance in this context, where intermediate D2D relays can be powered through harvested energy, enhancing both coverage and network reliability~\cite{Ashraf:2021}. 
However, practical implementation challenges exist. The nonlinear behavior of EH circuits, limitations in RF-to-DC conversion efficiency, and the trade-off between power and information decoding require careful transceiver and protocol design. Techniques like power splitting and time switching are commonly adopted to address the limitations of practical EH receivers in D2D networks~\cite{Ashraf:2021}.

\begin{figure}
   \centering
	\includegraphics[width=0.7\linewidth]{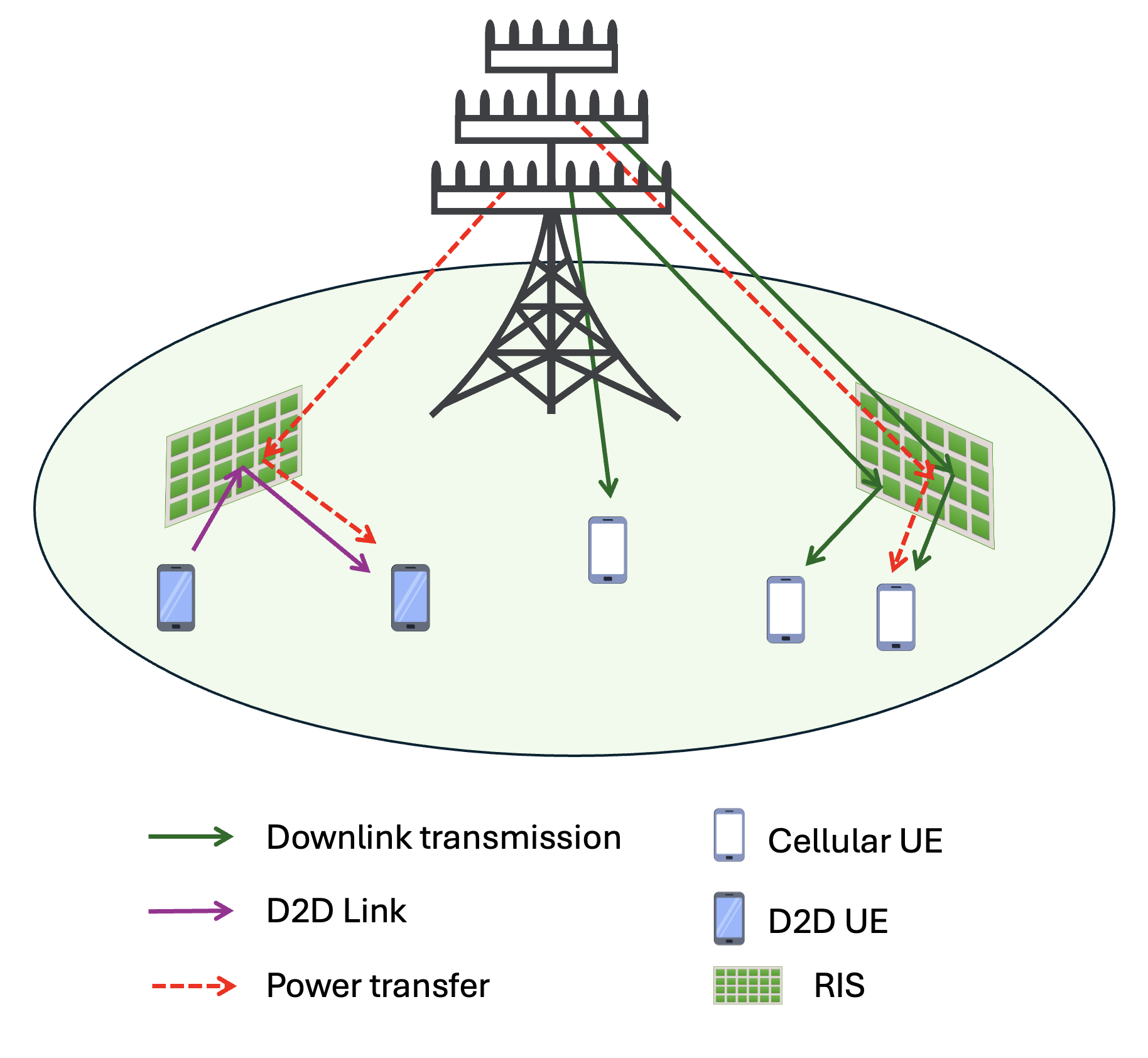}
	\caption{RIS-assisted massive MIMO D2D communication.}
	\label{fig:d2d_RIS}
\end{figure}

A comprehensive survey on SWIPT integrated with cooperative relaying systems, emphasizing their relevance for 5G and beyond wireless networks is provided by~\cite{Ashraf:2021}. The survey discusses both linear and nonlinear EH models for SWIPT under imperfect CSI and hardware impairments and explores various configurations including multi-relay and multi-antenna setups. The authors of~\cite{Shaik:2020} considered an HD EH based D2D-assisted MIMO communication system with imperfect CSI scenario over Nakagami-m channels. They showed how a small variation in the CSI correlation can deteriorate the performance of the system. A massive MIMO-assisted amplify-and-forward multi-pair relaying architecture was proposed by~\cite{Guo:2017} to enhance D2D communication performance in underlay settings. The proposed system simultaneously supports D2D communication and wireless EH by allowing D2D signals to be received and forwarded by a massive MIMO relay, while D2D users opportunistically perform information detection or EH from the relay's broadcast at the second hop. 

\subsection{Reconfigurable Intelligent Surfaces}

Reconfigurable Intelligent Surfaces (\gls{RIS}) is an emerging technology that enables dynamic control over the wireless environment through passive and software-controlled signal reflection. Unlike traditional transceiver-based systems, RIS consists of a large number of low-cost passive elements capable of independently adjusting the phase and/or amplitude of incident signals. This capability allows RIS to shape wireless channels in real time by intelligently directing signal paths, thereby enhancing signal strength, mitigating interference, and improving overall SE and EE~\cite{Wu:2020}.

Massive MIMO offers strong gains in spatial multiplexing and beamforming by equipping base stations with a large number of antennas. However, its performance can suffer in NLoS environments or when constrained by hardware challenges, such as high power consumption and increased complexity, particularly in mmWave frequencies. A RIS can help overcome these issues by passively shaping the wireless environment to create virtual LoS paths, improve channel quality, and fill coverage gaps. Unlike active components, RIS does not require additional RF chains, making it a more energy-efficient and cost-effective solution, especially in dense urban areas or at the edges of cells~\cite{Liu:2024_2}.

D2D communication further enriches this setting by enabling direct communication between UEs. When combined with RIS-assisted massive MIMO, D2D can help offload traffic from the core network and mitigate CCI. For example, RIS can be deployed near clusters of D2D users to facilitate local communication through intelligent reflection, while simultaneously suppressing interference toward the cellular users or BSs. Moreover, the integration of D2D and RIS in a massive MIMO framework provides more opportunities for exploiting spatial DoFs and user diversity. By coordinating the RIS configuration with D2D link scheduling and massive MIMO beamforming, the network can dynamically balance between direct transmission, reflected paths, and peer-to-peer communication, leading to significant gains in throughput, latency reduction, and network scalability~\cite{Fu:2021}. Fig.~\ref{fig:d2d_RIS} illustrates an RIS-assisted massive MIMO D2D communication network.

An interference alignment method for RIS-assisted MIMO D2D networks was proposed by~\cite{Fu:2021}, where the RIS assists the interference alignment in direct-link rank-deficient channels to reduce the CCI among multiple transceiver pairs, which leads to an increase in achievable DoFs. A comprehensive performance evaluation of RIS-assisted FD MIMO bidirectional D2D communications under realistic condition that include hardware impairments and residual self-interference was provided by~\cite{Bhowal:2022}.  In~\cite{Bhowal:2023}, a comprehensive model for realistic RIS-assisted FD MIMO D2D communication was developed using dual polarization to improve reliability and support polarization diversity. This approach is a robust and practical framework that incorporates hardware imperfections, polarization effects, and realistic indoor/outdoor channel conditions. A novel framework that integrates RIS and D2D communication into multigroup multicast MIMO systems was proposed by~\cite{Li:2024}. The authors extended RIS-based multicast beyond single-group scenarios and introduced D2D-assisted intra-group communication using a many-to-many resource reuse strategy to enhance SE. They also developed low-complexity optimization algorithms to jointly improve system coverage, group rate, and EE. Reference~\cite{Rao:2024} investigated secure RIS-aided D2D communication in cognitive cellular networks under energy constraints, where the D2D transmitter harvests energy via wireless power transfer. Another work~\cite{Liu:2024_2} studied RIS-assisted downlink cellular networks with spectrum-sharing D2D users and a multi-antenna BS employing beamforming and interference cancellation strategies under imperfect CSI by exploring ergodic achievable rate across various scenarios.

\section{Interplay between D2D and mmWave}
\label{sec:d2d_mmwave}

\subsection{Overview of mmWave Communication}

\begin{figure*}[t]
    \centering
	\includegraphics[width=\linewidth]{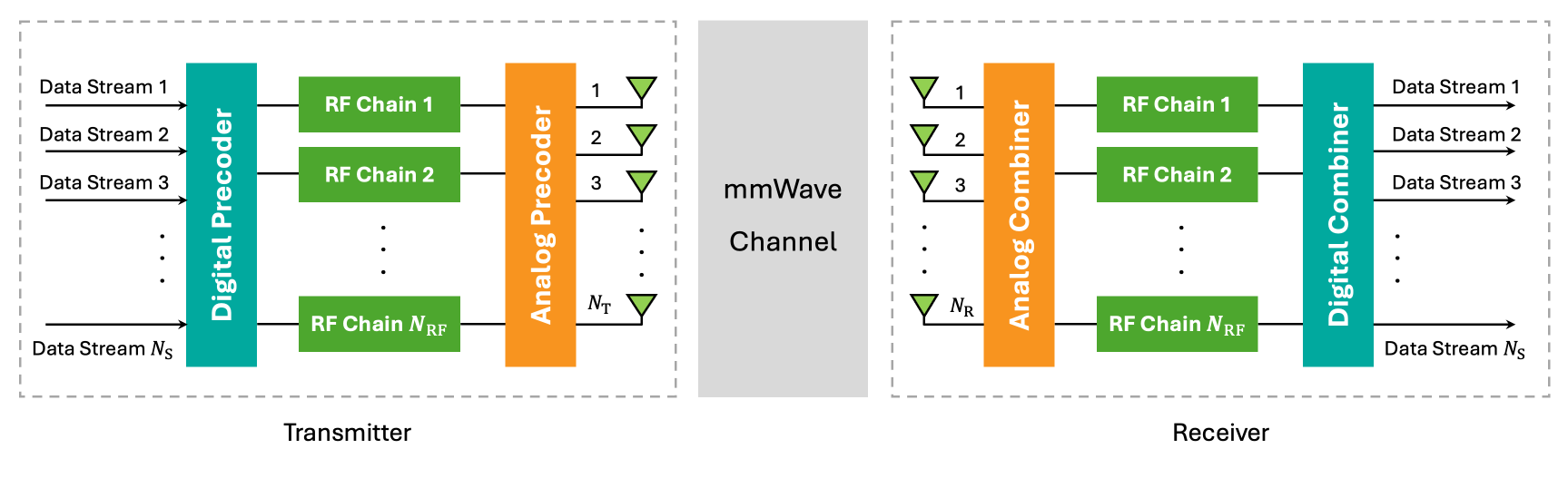}
	\caption{The structure of transceiver with hybrid beamforming in a mmWave massive MIMO system. }
	\label{fig:hybrid_precoder}
\end{figure*}

Millimeter wave (mmWave) communication operates within the 30--300~GHz frequency range and is a key enabler for 5G and beyond due to its potential to deliver ultra-high data rates, low latency, and large bandwidth. Unlike sub-6 GHz bands, mmWave offers abundant spectrum resources that support multi-gigabit-per-second (Gbps) throughput. This makes it well-suited for data-intensive applications such as augmented reality, high-definition video streaming, and vehicular networks~\cite{Qiao:2015}.

Recent progress in RF integrated circuit design has enabled the use of mmWave communication in outdoor environments. The short wavelength of mmWave signals make it feasible to pack a large number of antenna elements into a compact area. This capability supports massive MIMO systems, where highly directional beamforming can be used to transmit and receive signals. Such directionality improves spatial and beamforming gains and minimizes interference, which helps mitigate the signal degradation caused by the inherent high path loss of mmWave propagation~\cite{Turgut:2019}. 

Despite these advantages, mmWave signals are prone to several physical limitations, including high path loss, sensitivity to blockages (e.g., from buildings or even human bodies), and poor diffraction. These issues necessitate dense BS deployment, precise beam alignment, and the use of directional beams to maintain reliable connections. Although these techniques enhance signal coverage and reliability, they also introduce increased complexity and overhead, particularly in mobile or dynamic scenarios. Traditional digital beamforming is not well-suited for mmWave systems, as it requires a separate RF chain for each antenna, resulting in high cost and energy consumption. Conversely, analog beamforming is more power-efficient but limited to serving a single data stream via phase shifters~\cite{Solaiman:2021}. 

Hybrid beamforming offers a practical compromise by combining a low-dimensional digital baseband processor with an analog RF beamformer. This design reduces hardware complexity and power usage by relying on a smaller number of RF chains. At the same time, it supports multiple data streams through spatial multiplexing and delivers SE comparable to fully digital beamforming~\cite{Zhang:2020}. Fig.~\ref{fig:hybrid_precoder} illustrates a general hybrid beamformer for a transceiver of a mmWave massive MIMO system. 

\subsection{Integration of D2D with mmWave Communication}

\begin{figure}[t]
   \centering
	\includegraphics[width=0.7\linewidth]{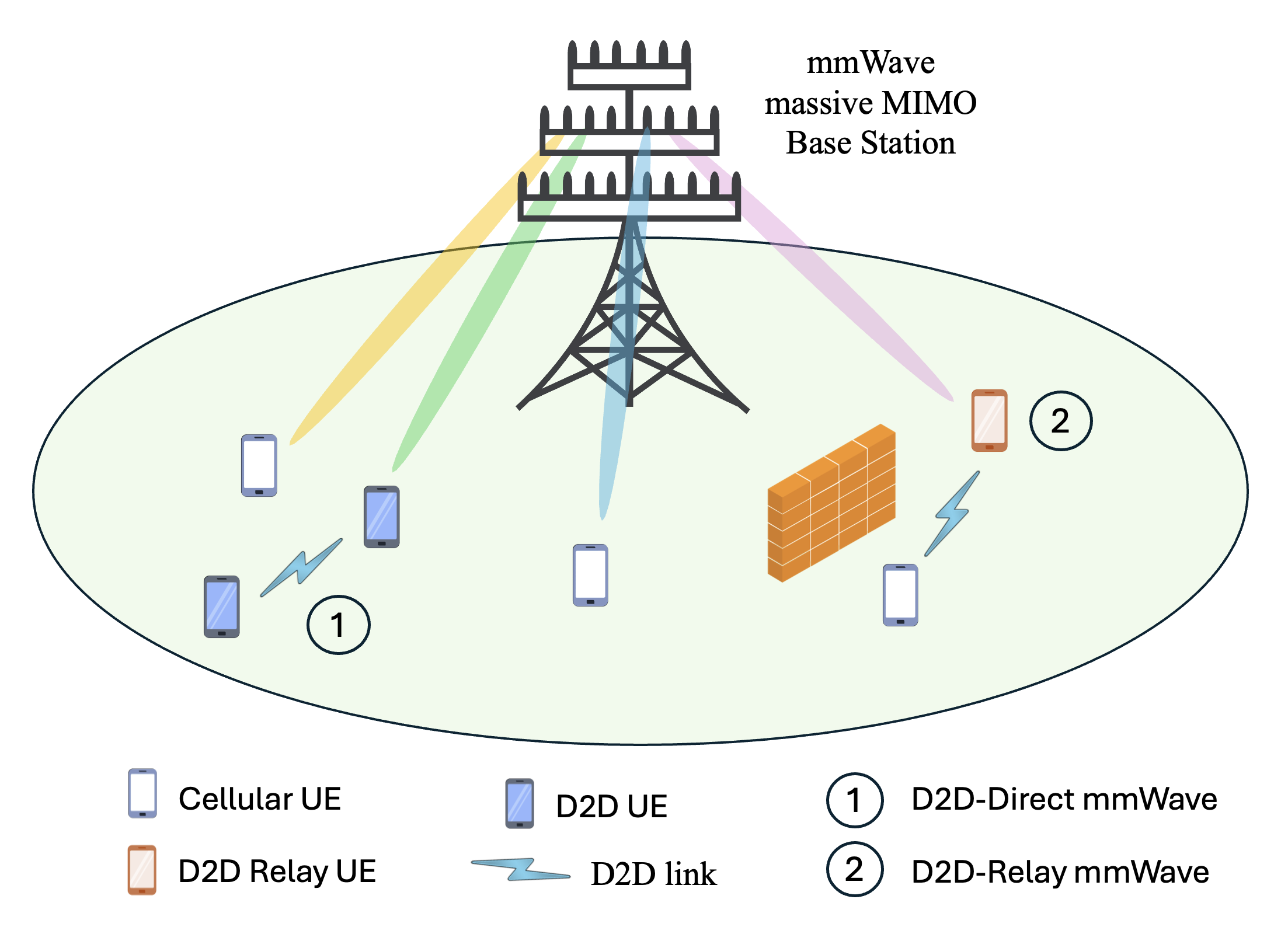}
	\caption{Modes of operation for D2D-enabled mmWave communications.}
	\label{fig:D2D_mmWave_modes}
\end{figure}

The integration of D2D communication with mmWave technology enhances 5G and beyond networks by combining 
proximity-based data exchange with high-throughput mmWave links. This integration can be realized through two primary modes: D2D-direct mode and D2D-relay mode. In the direct mode, devices communicate without any intermediate nodes and can operate autonomously or under network supervision. In the relay mode, D2D links act as supplementary hops that help maintain connectivity when direct paths are blocked due to obstacles. Fig.~\ref{fig:D2D_mmWave_modes} illustrates these modes for integration of mmWave-D2D in 5G cellular networks. 

\subsubsection{D2D-Direct mmWave Mode}

In conventional D2D-enabled cellular networks, managing CCI between D2D pairs and cellular users poses a major challenge. Communications at mmWave frequencies helps mitigate these issues through highly directional transmission enabled by large-scale antenna arrays. The resulting beamforming gain significantly reduces CCI, allowing multiple D2D pairs to reuse the same spectrum efficiently. This spatial reuse not only improves spectrum utilization but also reduces signaling and data overhead at the base stations. Moreover, since mmWave D2D links operate on separate high-frequency bands with minimal interference to sub-6 GHz cellular transmissions, they can support simultaneous D2D and cellular communication more effectively. These properties make mmWave D2D communication a promising solution for applications demanding high data rates and low latency, reinforcing its significance in 5G and future wireless systems~\cite{Deng:2017}.

\subsubsection{D2D-Relay mmWave Mode}

Due to their high susceptibility to blockage and limited diffraction, mmWave signals often suffer from link interruptions. D2D-relay communication addresses this issue by establishing multi-hop links through nearby devices, effectively bypassing obstacles and extending coverage. Selecting suitable relay nodes is essential and depends on factors such as the location and size of potential obstacles~\cite{Singh:2023}.

\subsection{Propagation Sensitivity and Blockage Modeling}

Signals in the mmWave regime are highly vulnerable to blockages due to their poor diffraction and penetration capabilities. This results in frequent link outages, especially in environments with dynamic or dense obstacles (e.g., buildings, vehicles, or pedestrians). For D2D communications, this sensitivity demands accurate modeling of obstacle locations and movement patterns. Existing solutions include using satellite imagery or learning based approaches such as occupancy grids, where historical link success/failure is used to infer blockage regions. Enhancing these models with spatial correlation (e.g., via Gaussian processes) can improve prediction accuracy and reduce training overhead~\cite{Sarkar:2020}.

Early studies adopted stochastic geometry frameworks using homogeneous PPPs to model the spatial distributions of BSs, UEs, and obstacles. For instance, \cite{Hourani:2016} employed a stochastic geometry approach to examine the EE of relay-assisted D2D systems by defining an energy saving zone where the relays significantly reduce energy consumption compared to conventional cellular transmission. Similarly, \cite{Jung:2016} analyzed connectivity performance by evaluating the probability of total connectivity and the average number of reliably connected devices, proposing a hybrid mode that dynamically switches between direct and relay-aided D2D links. These works often rely on simplifying assumptions such as an infinite number of users and propagation over an unbounded area, which facilitate the derivation of tractable expressions but fall short of capturing the spatial correlation effects present in dense scenarios.

To address the limitations of infinite-region models, \cite{Venugopal:2016} proposed a finite-region framework in which human bodies are treated as the dominant source of blockage. Their analysis computes the spatially averaged coverage probability and achievable rate by incorporating realistic path-loss and small-scale fading models. This approach more accurately reflects interference and blockage dynamics in crowded environments, where the same obstacle may simultaneously affect multiple D2D transmissions.

Despite their utility in average-case analysis, PPP-based models offer limited insight into link-specific performance due to the averaging over spatial realizations. To overcome this limitation, the concept of meta distribution has been introduced in the context of D2D-enabled mmWave systems. The meta distribution captures the distribution of the conditional success probability of individual links, providing a finer-grained perspective on network reliability. In this direction, \cite{Haenggi:2016} proposed a general framework for computing meta distributions, while \cite{Deng:2017} utilized it to evaluate how key parameters like D2D link distance and user density affect the signal-to-interference-plus-noise ratio (\gls{SINR}) and achievable rate. A more comprehensive study of link-level and network-level performance using meta distributions for D2D-underlaid mmWave cellular networks was presented in \cite{Wang_3:2021}. The authors in~\cite{Quan:2025} analyzed a mmWave D2D network tailored for URLLC using a stochastic geometry framework by incorporating multiple-antenna beamforming and modeling the reliability through the meta distribution of rate coverage probability.

\subsection{Beam Alignment and User Scheduling}

D2D links at mmWave require highly directional antennas for both transmission and reception. This necessitates precise beam alignment, which becomes increasingly difficult in the presence of user mobility or changing environments. Beam tracking errors can cause severe degradation in link quality or complete failure. Solutions such as fast beam training, predictive tracking using motion models, or reinforcement learning based beam selection 
with minimal signaling overhead have been proposed to address this challenge.

A common assumption in many meta distribution-based analyses is that of perfect beam alignment, where the maximum antenna array gain is always achieved by precisely steering the main lobe toward the desired receiver. In practice, beam misalignment is inevitable and can lead to severe performance degradation. To address this, \cite{Quan:2022} analyzed the impact of beam alignment errors by modeling them by a truncated Gaussian distribution. By integrating beam misalignment effects into the meta distribution framework, they demonstrated the significant sensitivity of coverage probability to alignment inaccuracies, which highlights the need for robust beam management in practical D2D-enabled mmWave deployments.

Another key limitation of many existing beamforming design approaches is their reliance on a predetermined set of scheduled users. In practical scenarios, particularly in ultra-dense D2D-enabled mmWave networks where the number of users requesting communication services often exceeds the number of simultaneously supportable links, beamforming alone is insufficient to achieve optimal system performance. To address this, user scheduling must be jointly considered alongside beamforming to effectively manage interference and enhance spatial reuse.

Several studies have explored joint beamforming and user scheduling strategies to optimize the performance of ultra-dense mmWave networks. For instance, \cite{He_2:2017} proposed a joint optimization framework for analog beam selection and user scheduling in a single-cell downlink multi-user multiple-input single-output (\gls{MISO}) system, utilizing limited CSI. In \cite{Niu:2018}, simultaneous transmission scheduling was investigated for mmWave small cells by leveraging D2D communication, concurrent transmissions, and multi-level beamforming codebooks to enhance SE. Moreover, \cite{Ye:2020} tackled the problem of jointly maximizing concurrent links and minimizing link length under Rayleigh fading and multi-user interference, presenting both centralized and distributed scheduling algorithms. The complexity of joint beam selection and link activation in ultra-dense D2D-enabled mmWave networks was analyzed in \cite{He:2022}, where the problem was shown to be NP-hard. To solve it efficiently, the authors proposed a deep learning based solution utilizing a graph neural network (\gls{GNN}) architecture.

Beyond scheduling and beam selection, energy efficiency and link reliability also present major challenges. In this context, \cite{Ma:2018} considered an FD relay-assisted D2D-enabled mmWave system and proposed a distributed relay selection algorithm based on matching theory. This approach aims to reduce the transmission power requirements of mobile devices while improving their achievable data rates—particularly important in scenarios with limited battery capacity and unstable D2D links.

\subsection{Transmit Antenna Selection}

Transmit antenna selection has gained considerable attention in mmWave communications, particularly for D2D systems. In this approach, the transmitter is equipped with multiple antennas, each designed to cover specific regions of the propagation space. By selecting the most suitable antenna, the system can direct a highly focused signal toward the intended receiver, improving link quality while minimizing interference. Unlike conventional systems, where the BS typically handles beamforming through costly hardware, D2D devices can instead utilize compact arrays of low-cost antennas with diverse directivity patterns. This enables high-quality, directional links without the need for complex beamforming architectures. However, to fully exploit this capability, an effective joint strategy for channel allocation and antenna selection is essential. This becomes particularly challenging in HetNets and in uplink scenarios, where interference, resource contention, and dynamic topology must be carefully managed.

Reference~\cite{Chen:2020} considered the joint optimization of antenna selection and analog combiner weights under practical hardware constraints which aims to enhance SE while reducing hardware complexity and power consumption. The authors of~\cite{Yazdani:2022} proposed a joint transmit antenna selection and resource allocation scheme for D2D-enabled mmWave networks where D2D transmitters are equipped with multiple directional antennas, and managed to obtain the globally optimal solution for the formulated problem by utilizing the generalized Bender’s decomposition algorithm, which converges in a finite number of iterations.  A low-complexity resource allocation scheme for uplink mmWave D2D communications using directional switchable antennas instead of conventional high-complexity beamforming was presented in~\cite{Monemi:2023}. The scheme focuses on a two-tier network where new low-priority D2D users reuse resources from existing cellular users without disrupting their QoS. The authors formulated a joint transmit antenna selection, power, and channel allocation problem only for new users, which significantly reduced computational overhead.  In~\cite{Akhoundzadeh:2024}, the authors proposed a joint optimization framework for antenna selection, spectrum assignment, and transmit power control in D2D-enabled mmWave underlay cellular networks. The objective is to maximize both system throughput and EE. Given the complexity and high dimensionality of the problem, a multi-agent distributed DRL approach is adopted, enabling decentralized decision-making and efficient resource allocation in dynamic network environments.

\subsection{Resource Allocation}

Efficient resource allocation in D2D-enabled mmWave networks is a complex problem shaped by the unique characteristics of high-frequency communication, such as directional beamforming and vulnerability to blockage. These factors, when combined with the proximity-aware nature of D2D links, demand advanced solutions that dynamically adapt to the spatial and temporal properties of the network.

A joint resource allocation and power control algorithm for hybrid mmWave-microwave networks with coexisting D2D communication is proposed in~\cite{Naqvi:2018} with a goal of maximizing the EE of cellular users while ensuring a minimum QoS for D2D users. In \cite{Hong:2020}, the resource allocation problem for D2D-enabled mmWave networks is addressed with the goal of maximizing the sum rate of D2D transmitters while managing interference to the base station. Subchannel allocation is performed using heuristic algorithms, while power allocation is handled through difference of convex programming.
In~\cite{Guo:2023}, the authors addressed resource allocation in D2D-enabled wireless networks for 5G mmWave and 6G THz scenarios, aiming to enhance spectrum utilization while ensuring QoS for both cellular and D2D users. A decentralized approach by combining DRL and federated learning is proposed to protect user privacy and reduce communication overhead under varying network conditions. The authors in~\cite{Dutta:2023} addressed the joint relay selection and channel allocation problem in D2D-enabled mmWave networks. To maximize the number of activated D2D links while meeting their data rate requirements, they formulated a stochastic integer program that accounts for user mobility and SINR variations caused by obstacles through probabilistic constraints, and a greedy approximation algorithm was formulated to solve the problem.

In D2D-enabled mmWave heterogeneous cellular networks, the challenges become more complex due to the coexistence of macro- and small-cell tiers, diverse propagation characteristics across frequency bands, and the need to coordinate resource use between cellular and D2D links. As a result, advanced algorithms are required to balance spectral efficiency, user fairness, and interference mitigation across microwave and mmWave bands in multi-cell environments. The authors of~\cite{Chen:2018} studied the resource allocation problem in D2D-enabled heterogeneous cellular networks and developed a coalition formation algorithm to address it. They extended their work to the multi-band scenario in~\cite{Chen:2019} by considering the distinct propagation characteristics and interference challenges of each band, and formulated an optimization problem to maximize the total system transmission rate.

\subsection{Relay Selection and Multi-Hop Path Formation}

In relay selection, intermediate UEs are chosen to forward data between a source and destination pair when a direct path is blocked or unreliable. In more obstructed or complex environments, multi-hop communication paths involving multiple relays can be formed to ensure connectivity and robustness. These techniques help improve signal quality, extend communication range, and enhance overall network throughput. However, selecting relays is non-trivial due to resource constraints, such as limited relay availability and the rule that a relay may only serve one communication 
session at a time~\cite{Johnston:2018}.

A further challenge arises from the dynamic nature of the environment where the selected relays themselves may become blocked over time due to moving obstacles, degrading the performance of active D2D pairs. In such cases, the BS must promptly select alternative relays to prevent unnecessary delay and energy waste. For effective relay management, the BS requires accurate CSI of D2D pairs, which can be reported along with other local parameters such as battery level, channel availability, and perceived throughput.
The unpredictability of link quality in such dynamic environments makes it difficult to model the system perfectly. As a result, online decision-making approaches become necessary. Partially Observable Markov Decision Processes (\gls{POMDP}s) are particularly suitable for modeling the uncertainties in D2D-enabled mmWave link quality caused by intermittent blockages. The BS can infer D2D link blockage by monitoring patterns such as successive packet losses and by deciding whether to persist with the current relay or switch to a new one. Since acknowledgment packets (ACKs) can also be lost due to blockage, the BS may rely on historical ACK statistics to improve its 
decision-making~\cite{Zhang:2023}.

A simultaneous prediction and training based framework for producing a more complete occupancy map is proposed by~\cite{Ganesan:2022}, which utilizes the information available from the nearby cells to deal with uncertainties come with each measurement and learn the relay selection policy while reducing the communication overhead. In~\cite{Singh:2023}, a sequential decision framework was developed by modeling the problem within a finite horizon POMDP framework and finding a stationary policy which prompt the UEs when to stop sending packets on the current link and explore new relay links based on the number of consecutive ACK failures. The derived policies capture the trade-off between delay due to packet loss and the cost of exploring a new relay link. The work in~\cite{Krishna:2025} explored the use of D2D relaying to boost uplink throughput in mmWave-based IoT networks, emphasizing relay placement and neighbor selection. It identifies an optimal coverage area that facilitates efficient relay-assisted transmission and proposes a geometry-driven relay placement approach that exploits this favorable region. The authors in~\cite{Gholizadeh:2025} examined relay selection and routing strategies for cluster-based multi-hop mmWave networks to address the limited range of mmWave signals due to blockage.

\section{Interplay between D2D and Multiple Access}
\label{sec:d2d_multiple_access}

\subsection{Overview of Multiple Access Mechanisms}

Multiple Access (\gls{MA}) mechanisms enable efficient spectrum sharing among multiple users in cellular networks. Traditional orthogonal multiple access (\gls{OMA}) techniques, such as time division multiple access (\gls{TDMA}) and orthogonal frequency division multiple access (\gls{OFDMA}), allocate distinct time-frequency resources to users, effectively eliminating interference but leading to suboptimal spectrum utilization. 

Non-orthogonal multiple access (\gls{NOMA}) techniques allow multiple users to be supported by the same resource block either by applying different unique, user-specific spreading sequences (i.e., code-domain NOMA) or by deploying superposition coding (\gls{SC}) at the transmitter and successive interference cancellation (\gls{SIC}) at the receiver (i.e., power-domain NOMA)~\cite{Liu_2:2017}. Despite its improved spectral efficiency and user connectivity, NOMA requires careful user clustering and power allocation to balance system performance and receiver complexity~\cite{Islam:2017}. 

Recently, rate-splitting multiple access (\gls{RSMA}) has emerged as a generalized and robust MA technique. RSMA splits each user's message into a common part (decoded by multiple users) and a private part (decoded individually), providing a powerful tool to manage interference adaptively~\cite{Clerckx:2023}. Unlike NOMA, which requires strict power domain separation, RSMA accommodates various channel conditions and user requirements with less stringent receiver design, making it a promising candidate for future wireless networks~\cite{Everett:2024}. Sparse code multiple access (\gls{SCMA})~\cite{Liu:2017_2} is also gaining attention for 5G and beyond systems, offering grant-free access with low signaling overhead, making it particularly relevant in dense networks with sporadic traffic patterns. 

\begin{figure}
    \centering
	\includegraphics[width=0.7\linewidth]{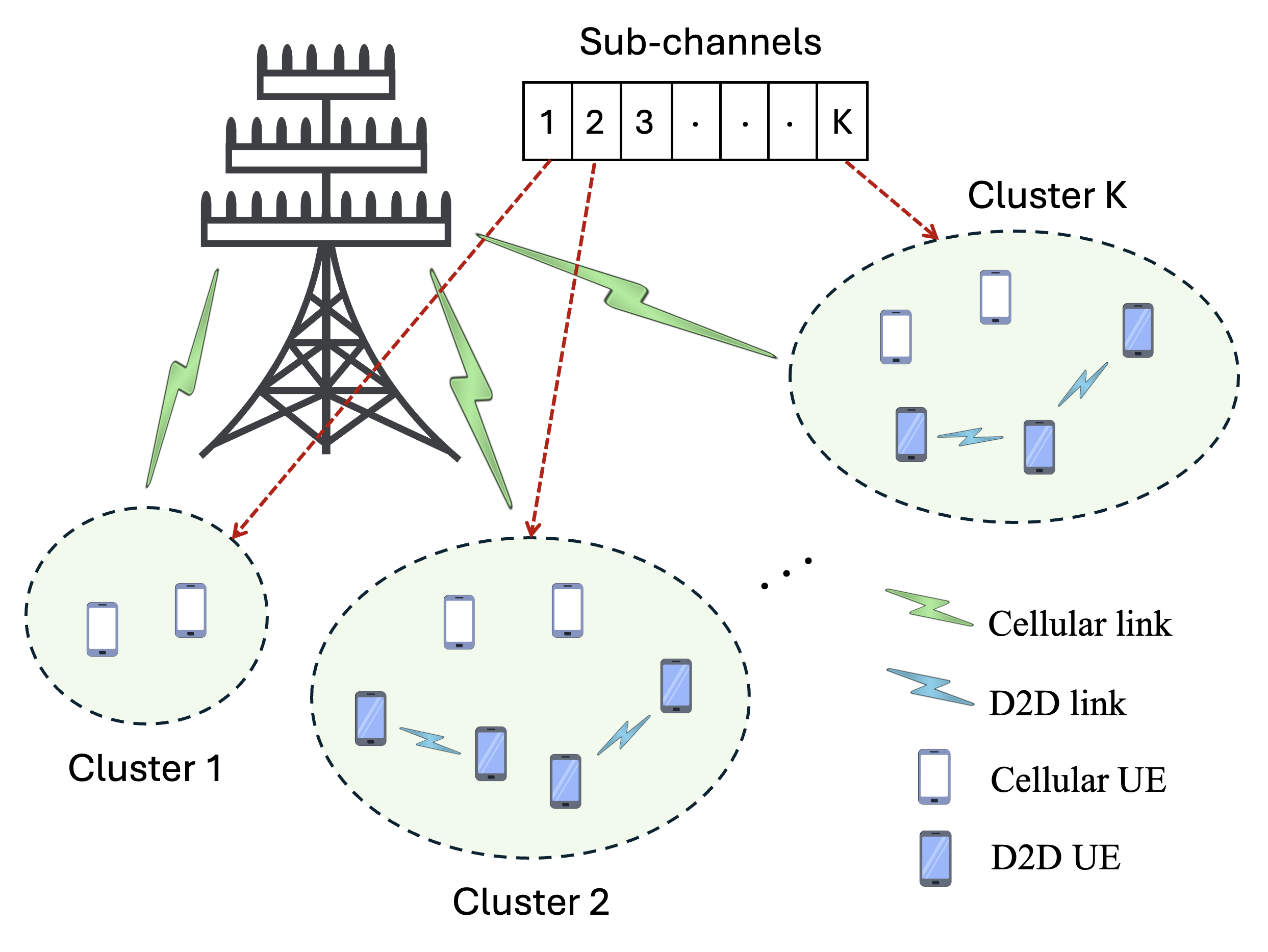}
	\caption{System model of D2D-enabled NOMA network.}
	\label{fig:d2d_noma}
\end{figure}

\subsection{Integration of D2D with Multiple Access Mechanisms}

Under OMA systems like OFDMA, D2D transmissions can be orthogonally scheduled to avoid interference, but with the cost of limiting the SE. The integration of NOMA with D2D provides a promising alternative. in NOMA-based D2D systems, multiple D2D pairs and cellular users can share the same RBs, with interference mitigated using power domain multiplexing and SIC, as shown in Fig.~\ref{fig:d2d_noma}. This enables a higher number of concurrent connections and improve spectrum utilization. However, challenges arise in clustering users, allocating power, and implementing SIC on resource-constrained D2D devices. 

Integration of RSMA into D2D communication, especially in scenarios involving channel uncertainty and dynamic interference pattern is another alternative. RSMA's capability to flexibly split message allows for enhanced interference management, making it suitable for relay-assisted or group-based D2D setups. Furthermore, RSMA reduces SIC dependency, offering a less-complex solution for low-power D2D devices. This is specifically beneficial in uplink scenarios, where interference at the BS is a critical issue. In IoT and mMTC contexts, D2D communication can be efficiently supported by SCMA, which provides contention-free access with minimal coordination.

To tackle the challenge of integrating D2D communication into NOMA-based 5G network, authors in~\cite{Dai:2019} introduced an innovative D2D mode by enabling power domain multiplexing between D2D pairs and cellular users while maintaining SIC feasibility at both the D2D receiver and BS. The joint clustering and power assignment problem in a NOMA-based cellular network with underlay D2D users is addressed in~\cite{Kazmi:2018}, where the user clustering is modeled as a matching game and power allocation solved via complementary geometric programming. In~\cite{Zhai:2019}, a joint user pairing, mode selection, and power control scheme for a NOMA-aided D2D system under decoding threshold constraints was introduced. An interference-aware user clustering method using graph theory, along with beamforming design and power allocation for a D2D-enabled MIMO-NOMA system, was proposed in~\cite{Solaiman:2021}.

The authors of~\cite{Alwani:2021} introduced a cooperative NOMA framework that enhances interference mitigation in MU-MISO-NOMA systems using beam-matching equalizers and D2D CSI sharing. By leveraging the concept of cooperative rate splitting,~\cite{Li:2017_2} addressed the challenge of mutual interference in D2D underlay cellular networks. Authors in~\cite{Ghosh:2025} provided a comprehensive performance analysis of an RSMA-based integrated sensing and communication (\gls{ISAC}) system under both infinite and finite blocklength regimes, considering imperfect CSI and SIC. By integrating NOMA and NR-U technologies into D2D network,~\cite{Sun:2020} proposed a joint optimization of licensed and unlicensed sub-channel allocation alongside a power control scheme, which aims to maximize SE in 5G heterogeneous ultra-dense networks. The work in~\cite{Agrawal:2022} explored the integration of cognitive radio principles into D2D communication within a unified cellular network, where all users are managed under the same infrastructure. A key assumption is that a full-duplex cellular user simultaneously supports its uplink to the BS and relays data for a D2D pair. 

Resource allocation in EH-powered uplink NOMA-based D2D communications was addressed in~\cite{Li:2022}. Recognizing the severe interference risks within D2D groups and between D2D and cellular users, especially under the constraints of fluctuating energy availability, the authors aimed to enhance EE through a distributed, game-theoretic framework. In~\cite{Kumar:2023}, the potential of multi-RIS-assisted D2D communication in a NOMA-enabled network to enhance SE for beyond-5G systems was explored. A comprehensive study of RIS-assisted NOMA frameworks for 6G wireless networks was given by~\cite{Sarkar:2024}. Reference~\cite{Vishnoi:2023} studied joint maximization of sum rate and fairness while considering power and resource constraints of both cellular users and D2D pairs. 
\section{Miscellaneous D2D Integrations}
\label{sec:d2d_misc}

\subsection{D2D communication with Localization}
\label{subsec:D2D_Localization}

Localization in wireless networks refers to determining the precise geographic position of a node, which supports applications such as emergency services, network optimization, and location-based services (\gls{LBS}) such as mapping, AR/VR, and vehicular positioning~\cite{Azeem:2023}. While satellite navigation remains the dominant technology due to its global coverage and high accuracy, it faces challenges in urban and indoor environments where signals may be obstructed~\cite{Peral:2018}. Wireless local area network (\gls{WLAN}) fingerprinting can achieve meter-level accuracy but requires maintaining extensive databases, and next-generation applications demand sub-meter accuracy~\cite{Peral:2018}. 

Cellular network localization, evolving alongside cellular standards from 1G to 5G, offers a
complementary solution. In particular, 5G NR enhances positioning through features like wide bandwidth, mmWave communication, massive MIMO, dense network deployment, and D2D communication. These advances support ubiquitous and high-speed positioning~\cite{Dammann:2015}. Reference~\cite{Peral:2018} studied the integration of cellular positioning into the cellular mobile radio standards and the development of location methods in different generations of cellular networks. A holistic overview of collaborative localization methods using D2D communication is provided in~\cite{Chukhno:2022}. Positioning techniques in 5G networks, with a focus on key enabling features such as mmWave, mMIMO, ultra-dense networks, D2D communications, and RIS along with the positioning accuracy enhancement techniques in industrial environments were investigated in~\cite{Muthineni:2023}. 

Cooperative localization via D2D allows devices to localize themselves by exchanging information directly, reducing reliance on BSs and improving accuracy and coverage, especially in dense heterogeneous networks~\cite{Chukhno:2022}. Recent studies demonstrated centimeter-level accuracy using mmWave D2D links and propose collaborative fingerprinting techniques enhanced by D2D communication for indoor positioning. The authors of~\cite{Cui:2016} investigated a real-time positioning scheme based on mmWave D2D links and claimed that their approach was able to provide up to centimeter level accuracy. Reference~\cite{Khandker:2019} introduced a D2D-based collaborative method for indoor positioning in which each UE obtains its estimated position using fingerprinting and refines it by cooperating with nearby devices through D2D links. 

\begin{figure}
    \centering
	\includegraphics[width=0.7\linewidth]{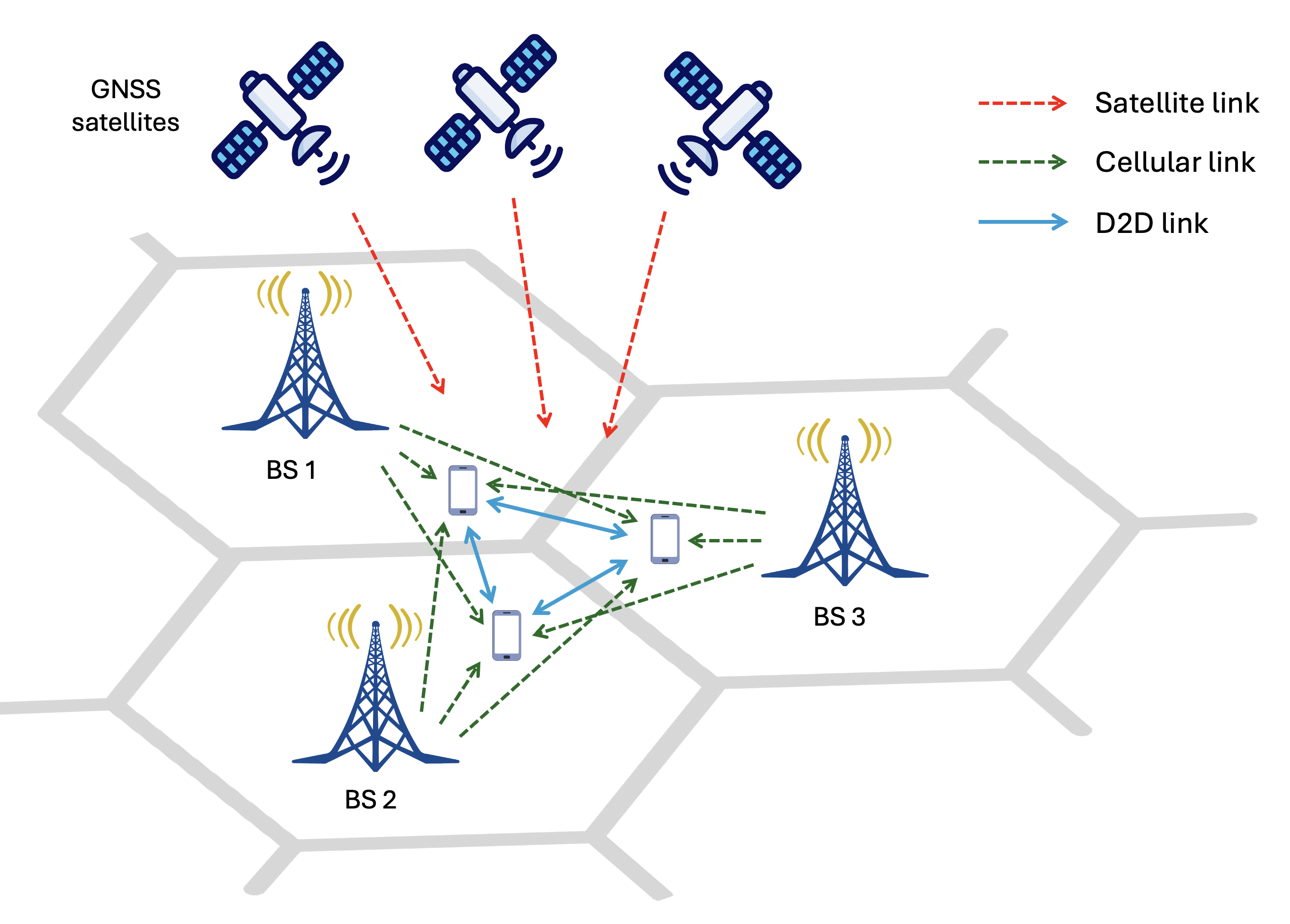}
	\caption{System model of GNSS/5G/D2D integrated positioning.}
	\label{fig:d2d_localization}
\end{figure}

A collaborative D2D-based localization method was proposed by~\cite{Raveneau:2017}, which aims to preserve the privacy of users by breaking the links between users and data. By leveraging matrix concatenation and multiplication, \cite{Liu:2022} devised two protocols for disguising the location and distance information into a random matrix while preserving their accuracy. Hybrid approaches integrating global navigation satellite system (\gls{GNSS}) with D2D measurements have shown significant improvements in accuracy~\cite{Yin:2018}, as illustrated in Fig.~\ref{fig:d2d_localization}. Other low-complexity methods use nearby device measurements to estimate AoA and enhance positioning in harsh channel conditions. For example, \cite{Sellami:2021} proposed a multi-stage low-complexity localization technique for UEs experiencing harsh channel conditions. A real-time collaborative localization scheme for D2D-enabled 5G network was proposed by~\cite{Famili:2023}. 

\subsection{D2D Communication over Unlicensed Spectrum}
\label{subsec:D2D_U}

D2D communication has emerged as a promising solution to offload traffic from cellular infrastructure. Licensed spectrum, though more reliable, is often insufficient to accommodate the growing demand for D2D communication, especially in dense deployments such as IoT networks~\cite{Le:2022}. In contrast, the unlicensed spectrum is more accessible for short-range D2D communication, as the higher path loss at these frequencies aligns well with the low transmit power constraints imposed by regulations. Given the proximity of transceivers in D2D scenarios, unlicensed spectrum is a practical choice for further boosting throughput~\cite{Liu:2016}.

The concept of using unlicensed spectrum to alleviate network congestion was first adopted by operators through the expansion of LTE into unlicensed spectrum~\cite{3GPP:2013}. Despite its promise, early offloading schemes often struggled with inefficiency and unreliable QoS, largely due to Wi-Fi's performance limitations and the lack of coordination between LTE and Wi-Fi networks~\cite{Zhang:2017}. These challenges are further intensified in 5G NR-Unlicensed (\gls{NR-U}) scenarios, where both cellular and Wi-Fi transmissions coexist in the same unlicensed bands. Here, maintaining QoS for latency sensitive traffic becomes especially critical~\cite{Fasihi:2024}. 

\begin{figure}
    \centering
	\includegraphics[width=0.8\linewidth]{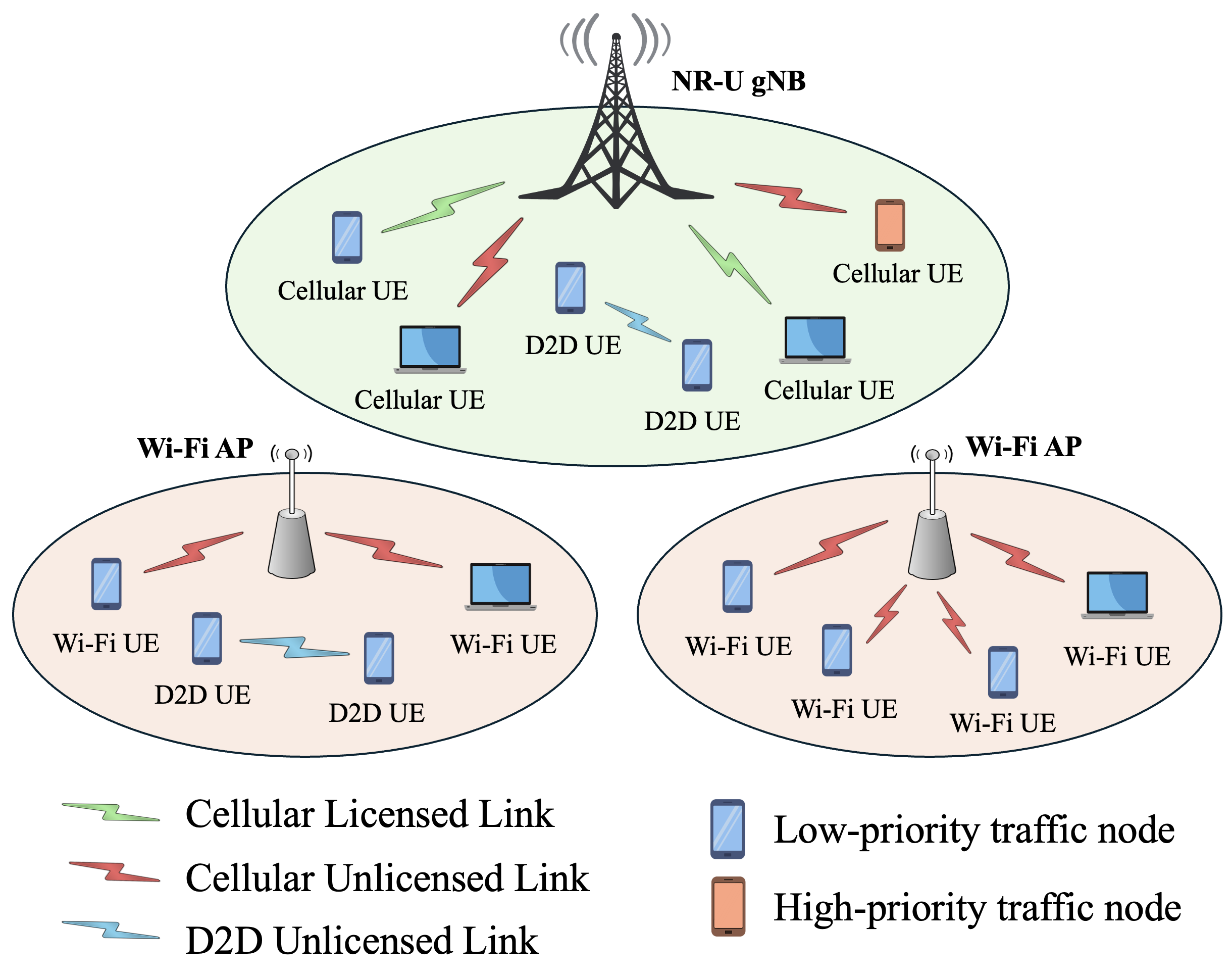}
	\caption{System model of coexistence of Wi-Fi, NR-U and D2D UEs on unlicensed spectrum.}
	\label{fig:d2d_unlicensed}
\end{figure}

Researchers introduced the concept of D2D communications over unlicensed spectrum, often referred to as D2D-Unlicensed (\gls{D2D-U}) communication~\cite{Wu:2019}, as shown in Fig.~\ref{fig:d2d_unlicensed}. Nonetheless, ensuring fair and efficient coexistence among heterogeneous systems (i.e., Wi-Fi, cellular, and D2D) in the unlicensed bands remains a significant challenge. The opportunistic nature of unlicensed spectrum access and mutual interference among these systems complicates harmonious operation~\cite{Zhang:2017}. Designing effective coexistence mechanisms for D2D-U requires addressing three primary issues. First, selecting the optimal communication mode, including the spectrum choice (licensed or unlicensed) and access mechanism (e.g., listen-before-talk or duty cycling), is essential. Second, joint resource allocation and power control across both spectrum types must be optimized. Finally, interference among D2D, cellular, and Wi-Fi users must be mitigated to prevent D2D-U transmissions from degrading the performance of incumbent systems~\cite{Liu:2016}. 

To address these issues, D2D-U networks must adapt to the heterogeneous nature of licensed and unlicensed access protocols. Recent studies have proposed intelligent coexistence management strategies aimed at dynamically tuning the coexistence parameters of incumbent technologies~\cite{Fasihi:2025}. When extended to D2D scenarios, such frameworks can help ensure that D2D links achieve acceptable QoS without compromising the performance of Wi-Fi or 5G NR-U systems.

\subsection{D2D Communication with SAGIN}

D2D communication was initially developed to facilitate direct data exchange between nearby devices without routing through centralized infrastructure and supporting LTE and 5G networks. Despite its advantages, contention-based D2D protocols can suffer from unbounded latency due to the lack of access guarantees, particularly under high traffic loads. While infrastructure-assisted D2D, where a terrestrial BS coordinates resource allocation to D2D links, offers more predictable and reliable performance, its dependence on fixed ground infrastructure makes it unsuitable for remote or infrastructure-sparse regions~\cite{Franke:2023}. 
To address these challenges, recent efforts have explored integrating non-terrestrial communication components into D2D frameworks. 

Advances in satellite technology, especially with low earth orbit (\gls{LEO}) constellations, have revolutionized the role of satellite communication in supporting direct and relayed D2D connectivity. Unlike traditional geostationary earth orbit (\gls{GEO}) and medium earth orbit (\gls{MEO}) systems, LEO satellites offer reduced latency, higher capacity, and global coverage, enabling effective D2D services in underserved or disconnected areas~\cite{Dou:2024}. These advancements are further complemented by the rise of the satellite-air-ground integrated network (\gls{SAGIN}), which seamlessly combines spaceborne, airborne, and terrestrial network elements into a unified architecture. SAGIN enhances the flexibility, resilience, and scalability of D2D systems by dynamically leveraging available resources across multiple layers of the communication ecosystem~\cite{Abuali:2023}.

Enabling efficient D2D communication within SAGIN, however, introduces significant technical challenges due to heterogeneous network elements, dynamic topologies, and varying link conditions. Adaptive and intelligent control strategies are essential to address issues such as mobility management, resource allocation, energy efficiency, and secure data exchange. Emerging approaches leverage federated learning to enable decentralized, privacy-aware optimization of D2D transmissions across SAGIN layers~\cite{You:2023}. In parallel, blockchain-based frameworks are being investigated for ensuring secure and verifiable interactions among distributed network entities. Additionally, sustainable operation is becoming increasingly important, prompting the integration of low-carbon computing techniques to reduce the environmental footprint of globally connected networks~\cite{Wang:2024}. These innovations collectively push the boundaries of 6G connectivity, enabling robust, secure, and universally accessible D2D services within the broader SAGIN paradigm. 

\subsection{D2D communication with ISAC}

With the rapid expansion of IoT devices and the proliferation of applications requiring simultaneous high-speed communication and high-precision sensing (e.g., intelligent transportation, smart manufacturing, and human-computer interaction), spectrum resource congestion has become increasingly pronounced. Integrated sensing and communication (ISAC) has emerged as a promising technology in 6G networks, enabling the dual functions of communication and sensing over the same wireless infrastructure and spectral resources. ISAC systems, supported by advancements in MIMO technologies, signal processing, and hardware design, improve both SE and EE by facilitating shared usage of communication and sensing capabilities~\cite{Dou:2024}.

In ISAC systems, the BS typically handles both communication and radar sensing, with MIMO beamforming playing a vital role in balancing these dual functions. Most of the research has been focused on beamforming at the transmitter side~\cite{Liu:2025}. However, achieving full-duplex ISAC operation introduces the challenge of self-interference, which must be mitigated effectively. To enhance ISAC scalability and offload BS resources, D2D communication offers a promising solution. D2D enables direct communication between nearby devices, reducing latency, increasing spectral reuses, and alleviating the load on base stations, which is especially beneficial in the complex environment of ISAC~\cite{Jiang:2025}.

On the other hand, D2D communication introduces interference management and complexity challenges, especially when multiple D2D pairs share spectrum. Although various power control algorithms have been introduced recently to optimize power allocation, many of them assume homogeneous D2D networks and do not fully address the complexity of cellular-underlaid or heterogeneous scenarios~\cite{Xue:2024}. Continued research into join transceiver design, full-duplex operation, and intelligent distributed control will be essential for realizing the full potential of D2D-aided ISAC networks. 
\section{Future Research Challenges in D2D-Integrated 5G/6G Networks}
\label{sec:future_research}

While the integration of D2D communication within 5G and emerging 6G networks has been well-studied in the literature, numerous open challenges must still be addressed. The complex interplay between D2D communication and enabling technologies presents new technical, architectural, and operational issues. The following challenges highlight critical areas for future research.

\subsection{Standardization and Interoperability}

Currently, D2D communication lacks consistent global standards across cellular, satellite, and unlicensed bands. This creates interoperability issues between different manufacturers and network provides. Without unified protocols, it becomes difficult to deploy D2D in large-scale, multi-vendor, and multi-technology networks. Standardization is also needed to ensure security, resource sharing, and efficient spectrum usage across diverse use cases. 

\subsection{Efficient Mode Selection}

D2D can operate in various modes such as direct mode, relay-mode, or cellular mode. Selecting the appropriate mode based on network conditions, user mobility, and application requirements is a complex problem. This becomes even harder in heterogeneous networks that include mmWave links, UAVs, or satellite nodes, where link reliability and latency vary widely. Future work should investigate AI-based mode selection frameworks that adapt in real time to changing network conditions and user requirements.

\subsection{Handling Channel Feedback in Massive MIMO}

Massive MIMO systems can significantly boost D2D throughput and SE through spatial multiplexing. However, they require timely and accurate CSI from each device. Gathering and updating this information, especially in dynamic and dense environments leads to signaling overhead and increased energy consumption. Research is needed on lightweight, predictive CSI acquisition methods and beamforming strategies that reduce feedback overhead while maintaining performance.

\subsection{Overcoming Blockage in mmWave}

The mmWave spectrum offers large bandwidths for high-speed D2D links, but is highly susceptible to physical blockage from buildings, vehicles, or even human bodies. Maintaining a reliable connection in such conditions requires fast beam alignment, backup paths, or relay nodes. Developing robust beam-tracking and prediction methods is essential for practical mmWave D2D deployments.

\subsection{Managing Interference in Multi-Access Networks}

D2D must coexist with multiple access schemes like OMA, NOMA, and RSMA, each with distinct interference characteristics. Coordinating D2D transmissions in such environments without degrading performance is difficult. Future efforts will need to include developing interference-aware resource allocation techniques and cross-layer coordination mechanisms that adapt to different access schemes.

\subsection{Sharing Spectrum in an Unlicensed Spectrum}

Using D2D in unlicensed bands can ease congestion in the licensed bands but requires effective coexistence with existing users like Wi-Fi. Techniques like listen-before-talk and spectrum sensing must be adapted for D2D's low-latency nature. Future studies should optimize spectrum sharing protocols and explore cooperative sensing approaches for dynamic, fair access in the unlicensed bands. 

\subsection{Integrating D2D into SAGIN}

Integrating D2D into SAGIN can provide robust coverage in remote or emergency scenarios. However, D2D links must adapt to latency, handover, and mobility issues unique to satellite and aerial segments. Future research should aim at developing adaptive D2D protocols that consider orbital dynamics, propagation delays, and cross-domain handoffs for seamless SAGIN integration.

\subsection{Combining D2D with Sensing in ISAC}

ISAC enables devices to sense the environment while communicating, opening up applications like simultaneous object detection and data transfer. However, balancing sensing accuracy with communication reliability is challenging. Research should focus on resource optimization and scheduling methods that enable efficient trade-offs between sensing and communication tasks in D2D-ISAC systems.

\section{Conclusions}
\label{sec:conclusions}

D2D communication has emerged as an essential component in the evolution of 5G and future 6G networks, offering significant improvements in spectral efficiency, latency, energy consumption, and network capacity. This paper has explored the architectural foundations of D2D communication and examined its convergence with a range of 5G/6G enabling technologies. Our findings highlight that while D2D significantly improves network performance and enables new services, its effective deployment requires careful design choices. Furthermore, standardization gaps and the complexity of coordinating D2D across different technologies and spectrum bands must be addressed to achieve 
wide-scale adoption. 

Looking forward, D2D communication is expected to play a major role in realizing the goals of 6G, particularly in supporting ultra-reliable low-latency communication, pervasive connectivity, and intelligent network services. Continued research is needed to develop adaptive, secure, and scalable D2D solutions that can seamlessly integrate with emerging network paradigms. 

\printglossary[type=main,nonumberlist]

\bibliographystyle{IEEEtran}
\bibliography{IEEEabrv,main}

\end{document}